\documentclass[10pt]{iopart}
\usepackage{anysize}
\usepackage{float}
\usepackage{wrapfig}
\usepackage[margin=2.2cm]{geometry}
\usepackage{graphicx,subfigure,wrapfig}
\usepackage{amssymb}
\usepackage{physics}
\usepackage{mathtools}
\usepackage{cancel} 
\usepackage[cal=boondoxo]{mathalfa}

\usepackage{bm}
\usepackage[usenames, dvipsnames]{xcolor} 
\usepackage{hhline}

\makeatletter
\newsavebox{\@brx}
\newcommand{\llangle}[1][]{\savebox{\@brx}{\(\m@th{#1\langle}\)}%
  \mathopen{\copy\@brx\kern-0.5\wd\@brx\usebox{\@brx}}}
\newcommand{\rrangle}[1][]{\savebox{\@brx}{\(\m@th{#1\rangle}\)}%
  \mathclose{\copy\@brx\kern-0.5\wd\@brx\usebox{\@brx}}}
\makeatother

\usepackage{etoolbox}
\makeatletter
\newrobustcmd{\fixappendix}{%
  \patchcmd{\l@section}{1.5em}{7em}{}{}%
  \patchcmd{\l@subsection}{2.3em}{7em}{}{}%
}
\makeatother

\usepackage{graphicx}
\usepackage[breaklinks,colorlinks = true,linkcolor = red,urlcolor=blue,citecolor=WildStrawberry]{hyperref}
\usepackage{ulem}
 \usepackage[mathscr]{eucal}
\expandafter\let\csname equation*\endcsname\relax
\expandafter\let\csname endequation*\endcsname\relax
\usepackage{amsmath}
\usepackage{psfrag,latexsym}
\usepackage{epstopdf}
\usepackage[ascii]{inputenc}
\usepackage{color}
\definecolor{dgreen}{rgb}{0,0.7,0}

\def\brvr{\bar{\varrho}}
\def\brr{\bar{\rho}}
\def\brf{\bar{\fn}}
\def\mtx{\mathtt{x}}
\def\mtt{\mathtt{t}}

\def\mff{\mathfrak{f}}
\def\fn{\mc{f}}
\def\hn{\mc{h}}
\def\Fn{F}
\def\gn{\mc{g}}
\newcommand{\mc}[1]{\mathcal{#1}}  
\newcommand{\msc}[1]{\mathscr{#1}}  
\newcommand{\hmsc}[1]{\hat{\mathscr{#1}}} 
 
\newcommand{\hmc}[1]{\hat{\mathcal{#1}}} 
\newcommand{\mb}[1]{\mathbb{#1}}  
\newcommand{\mf}[1]{\mathfrak{#1}} 
\newcommand{\mt}[1]{\mathtt{#1}} 
\newcommand{\bmf}[1]{\bar{\mathfrak{#1}}} 

\begin{document}
\title{{Ballistic} macroscopic fluctuation theory of correlations in hard-rod gas}
\author{Anupam Kundu}
\address{International Centre for Theoretical Sciences, TIFR, Bengaluru -- 560089, India}
\ead{anupam.kundu@icts.res.in}
\begin{abstract}
Recently, a theoretical framework known as {\it ballistic macroscopic fluctuation theory} has been developed to study large-scale fluctuations and correlations in many-body systems exhibiting ballistic transport. In this paper, we review this theory in the context of a one-dimensional gas of hard rods. The initial configurations of the rods are sampled from a probability distribution characterized by slowly varying conserved density profiles across space.
Beginning from a microscopic description, we first formulate the macroscopic fluctuation theory in terms of the phase-space density of quasiparticles. In the second part, we apply this framework to compute the two-point, two-time correlation functions of the conserved densities in the Euler scaling limit. We derive an explicit expression for the correlation function which not only reveals its inherent symmetries, but is also straightforward to evaluate numerically for a given initial state. Our results also recover known expressions for space-time correlations in equilibrium for the hard-rod gas.
\end{abstract}
\maketitle


\tableofcontents

\section{Introduction} 
Understanding non-equilibrium dynamics in many-particle systems is a fundamental problem in statistical physics \cite{polkovnikov2011colloquium, eisert2015quantum}. Often, the focus is on the large-scale evolution of macroscopic quantities, such as the densities of conserved quantities, starting from non-equilibrium states. Studying such evolution from a microscopic perspective is challenging, except in cases where particles (or degrees of freedom) do not interact. In systems with short-range interactions, hydrodynamic (HD) theory offers a universal framework \cite{spohn2012large,  spohn2024hydrodynamic, doyon2020lecture}. This theory is based on the assumptions of local equilibration and enables one to write  evolution equations for conserved densities, which form closed sets of differential equations \cite{landau1987fluid, doyon2020lecture, denardis2023hydrodynamic}.

At the largest space-time scales, HD theory yields the Euler equations, which describe ballistic evolution. To capture features in density profiles at smaller, sub-ballistic scales, dissipation terms - often introduced phenomenologically, such as viscosity and Fourier terms - are added, as in the Navier-Stokes equations for simple fluids \cite{hansen2013theory}. Although HD theory is a classical framework, it has been remarkably successful empirically across diverse contexts, including normal fluids \cite{martin1972unified}, magnetic liquids \cite{mryglod1996hydrodynamic}, ultra-cold atoms \cite{sommer2011universal}, plasma \cite{moisan2012hydrodynamic}, energy transport in low dimension \cite{spohn2014nonlinear, lepri1997heat} and active matter \cite{julicher2018hydrodynamic}. However, a universally accepted derivation of HD equations from microscopic dynamics remains elusive.

In the past decade, significant advances have been made in hydrodynamic theory, extending the hydrodynamic approximation to integrable systems that possess an infinite number of conserved quantities. This extension, known as generalized hydrodynamics (GHD), describes the evolution of these conserved densities, contrasting with the finite set considered in conventional hydrodynamics \cite{castro2016emergent, bertini2016transport, bastianello2022introduction, doyon2025generalised,  spohn2018interacting, doyon2019generalized, spohn2021hydrodynamic}. The GHD equations are elegantly formulated in terms of the evolution of stable quasiparticle densities \cite{doyon2020lecture}. While integrability is a fine-tuned property, many systems near integrability exhibit striking effects on large-scale relaxation dynamics. This has spurred intense interest in exploring GHD, which has proven highly successful in understanding ballistic-scale motion in many systems \cite{kinoshita2006quantum, caux2019hydrodynamics, malvania2021generalized, alba2021generalized}.

Another key objective in the statistical mechanics of many-particle systems is to understand the structure of correlations - both static and dynamic - and the fluctuations of local observables. This requires moving beyond the description of average conserved densities and associated average fluxes to develop a statistical theory of fluctuations on macroscopic scales. Originally, such an attempt was made by Landau and Lifshitz  by looking at the relaxation of a small perturbation in the initial stationary state  which can also be formulated alternatively by including a stochastic forcing to each dissipative flux \cite{ landau2013statistical, landau1992hydrodynamic, de2013non, de2006hydrodynamic}. 
This approach has recently been extended to compute equilibrium space-time correlations in the Toda system \cite{mazzuca2023equilibrium, spohn2020ballistic}. However, such a procedure fails to describe anomalous energy transport in low-dimensional systems \cite{alder1970decay,forster1977large, van2012exact} for which a nonlinear extension of the linear fluctuating HD theory has been introduced \cite{spohn2014nonlinear, spohn2016fluctuating}.  In the context of diffusive hydrodynamics, macroscopic fluctuation theory (MFT) provides a universal approach \cite{bertini2002macroscopic}. MFT, rooted in the large deviation theory, characterizes the probabilities of density and current fluctuations at diffusive scales \cite{bertini2015macroscopic}.

Recently, a similar large deviation framework for integrable systems has been developed, known as ballistic macroscopic fluctuation theory (BMFT) \cite{doyon2023emergence, doyon2023ballistic}. BMFT describes density and current fluctuations at ballistic scales and is broadly applicable, even to systems with a small number of conserved quantities that exhibit ballistic transport.
 BMFT is grounded in the principle of local relaxation, which essentially says that fluctuations of any local observable on the largest space-time scale are determined by the fluctuations of the conserved densities carried coherently {from the initial state}. 

One of the most {striking predictions}s of BMFT is the emergence of long-range correlations in systems with short-range interactions \cite{doyon2023emergence, doyon2023ballistic}. These correlations develop over long times in non-stationary states, even when starting from initial configurations with only short-range correlations. As an intriguing consequence of these long-range correlations, it has been shown that the standard Navier-Stokes correction to hard-rod dynamics remains valid \cite{hubner2024diffusive,hubner2025diffusive} only for a short initial period of time \cite{boldrighini1997one}. Using BMFT, these long-range correlations have been analytically computed, demonstrated, and numerically verified for hard rods in one dimension \cite{doyon2023ballistic, hubner2024diffusive}. A hard-rod gas in one dimension is a simple, yet non-trivial interacting integrable system for which many features of GHD and BMFT can be explicitly illustrated by performing both microscopic and macroscopic calculations \cite{doyon2017dynamics, singh2024thermalization, powdel2024conserved}.   

The goal of this paper is to review BMFT in the context of hard rods and to present an alternative description of BMFT in terms of the phase-space density of quasiparticles and, finally, to apply this theory to compute the unequal space-time correlation of conserved densities.
A collection of hard rods was first considered by L. Tonks in 1930 as the simplest model of an interacting gas that exhibits thermodynamic behavior distinct from ideal gas laws \cite{tonks1936complete}. 
Due to its solvable structure, this model and its variants have been widely used in equilibrium physics of simple fluids to compute and understand various thermodynamic quantities such as density fluctuations, pair correlation functions, and distribution functions \cite{tonks1936complete, robledo1986distribution, salsburg1953molecular, sells1953pair, koppel1963partition}. Later microscopic approaches were also used to study time evolution \cite{lebowitz1968time,bernstein1988expansion, jepsen1965dynamics, valleau1970time}. The hydrodynamic equation for the phase-space density was first derived by Percus from the Boltzmann equation \cite{percus1969exact} to study the dynamics of the gas on a macroscopic scale.

In a hard-rod gas, the particles undergo elastic collisions and are not allowed to approach each other closer than a distance
$a$, which can be interpreted as the interaction range or the rod length. Due to elastic collisions, the particles exchange momenta upon interaction. Several equilibrium properties, such as density fluctuations, pair correlation functions, and distribution functions, have been analytically computed for this model. As an artifact of integrability, the dynamics of a gas of
$N$ hard rods can be mapped to that of $N$ hard-point particles and vice versa. This mapping was previously used in several studies \cite{lebowitz1968time,percus1969exact} and has since been exploited to study non-equilibrium dynamics \cite{bernstein1988expansion}. With the advent of GHD, interest in this model has increased as a platform to demonstrate and test the predictions of both GHD and BMFT.

In this paper, we exploit the mapping to the hard-point gas to solve the BMFT equations and use these solutions to compute two-point, two-time correlations of conserved densities. We provide explicit expressions that are easy to evaluate numerically. Our results reproduce previous findings for the two-point correlation in hard-rod gas derived from Landau-Lifshitz theory in equilibrium and from a previous account of BMFT in the non-equilibrium initial state.

The paper is organized as follows. In section \ref{BMFT-hr} we start by {providing a microscopic description} of the hard-rod gas, which is followed by a development of the BMFT in terms of the phase-space density (PSD) of hard-rod quasiparticles. The BMFT equations are derived through a saddle-point method. We end this section by discussing how one can use this theory to compute the unequal space-time correlation of the PSD of hard rods. In the next section \ref{BMFT-hp}, we provide an equivalent formulation of {BMFT of the hard-rod gas} in terms of the PSD of a hard-point gas using an exact mapping of the microscopic configurations. Solutions of the BMFT equations are obtained in hard-point picture sec.~\ref{sec:den-corr}. We also show in \ref{derv:sad-eq-hr} that the same solution can also be obtained by solving the equations in the hard-rod picture. These solutions are finally used in sec.~\ref{sec:eva-corr} to compute the space-time correlations. We present computation of equilibrium space-time correlations in sec.~\ref{sec:unequal-C} whereas explicit expression of unequal space-time correlation of mass densities in the non-equilibrium state is presented in sec.~\ref{sec:corr-mass-den}. In sec.~\ref{conclusion}, we conclude our paper with possible future extensions. To enhance the readability of the paper, we have placed a substantial amount of calculations  in several appendices.

\begin{figure}[t]
\centering
\includegraphics[width=6.5in]{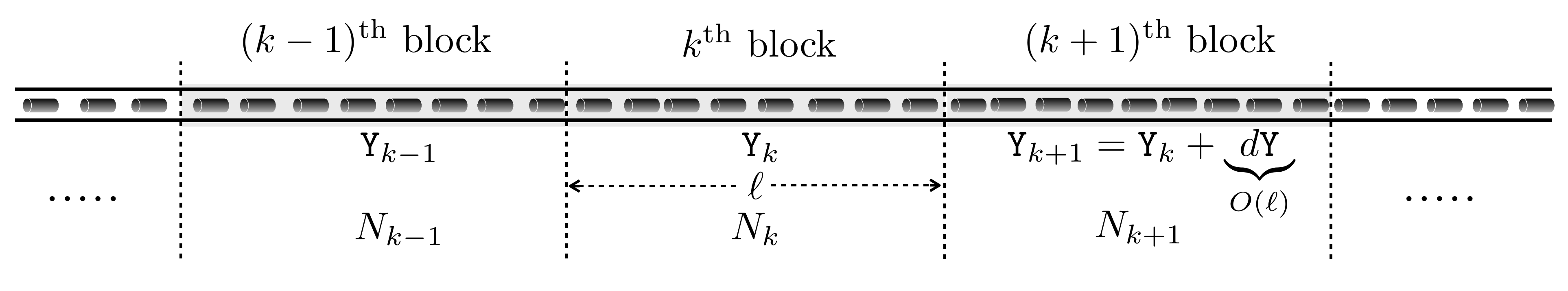} 
\caption{Schematic diagram of a gas of $N$ hard rods, each of length $a$ and unit mass in one dimension. The space of the system is divided into blocks of size $\sim \ell$ and labelled by $\mt{y}_k=k\ell$.  The number of rods also gets distributed among these blocks such that $\sum_kN_k=N$. We assume the following hierarchy of the length scales $a \ll \ell \ll Na$. We will eventuallyconsider the limit $N \to \infty$ and $\ell \to \infty$ however with $\ell/N \to 0$.}
\label{schema}
\end{figure}

\section{BMFT for hard-rod gas}
\label{BMFT-hr}
We consider $N$ hard rods, each of unit mass and length $a$, moving in a straight line. The positions and momenta of the rods are denoted by $\{\mtx_i,v_i\}$ for $i=1,2,...,N$. Since the mass of each particle is unity, the momentum of each particle is its velocity. These rods move ballistically and, when they meet, they collide elastically. At each collision, they exchange their momenta. 
Initially, the rods are placed inside a region of length $L$ and eventually we take the thermodynamic limit $N \to \infty$, $L\to \infty$ making the initial density profile a finite-valued function $\varrho_0(\mtx)$ everywhere. The dynamics of this system is integrable in the sense that it has $N$ locally conserved quantities $Q_\nu$, which can be chosen as the sum of the moments of the individual velocities. 
\begin{align}
Q_\nu=\sum_{i=1}^Nv_i^\nu, ~~\nu=1,2,...,N. \label{cons-Q}
\end{align}
We consider the following joint probability distribution from which the initial positions and velocities of the rods are chosen:
\begin{align}
\mathbb{P}_{\rm r}(\{\mtx_i,v_i\})=\frac{1}{Z_r}\prod_{i=1}^N\Psi(\mtx_i,v_i)\prod_{i=1}^{N-1}\Theta(\mtx_{i+1}-\mtx_i-a), \label{eq:mbP_r}
\end{align}
where $\Theta(\mtx)$ is the Heaviside theta function,  $Z_{\rm r}$ is the normalization factor, and $\int d\mtx \int dv~\Psi(\mtx,v)=1$. We assume that the initial state slowly varies over space, characterized by the following choice. 
\begin{align}
\Psi(\mtx,v)=e^{-\psi(\mtx/\ell,v)},  \label{Psi(x,v)}
\end{align}
with $\psi(x,v)$ being some suitable function that goes to infinity for $|x| \to \infty$ and $|v| \to \infty$. The parameter $\ell$ represents the length scale of the variation, which we assume to be large. For large $N$ and $\ell$, keeping $a \ll \ell \ll Na$, the partition function $Z_{\rm r}$ can be written as integrals over slowly varying fields.  To see that, we divide the space into regions of size $\sim \ell$ labeled by $\mt{y}_k=k\ell$ (see Fig.~\ref{schema}).  {These blocks are also called `fluid cell' in the literature \cite{doyon2023ballistic}.}  Breaking the integral into blocks of sizes $\ell$, the partition function $Z_{\rm r}$ can be approximated by  
\begin{align}
Z_{\rm r}\approx \sum_{\{N_k\}}~\delta_{\sum_kN_k,N}\prod_k\int_{\Box_\ell^k} \prod_jd\mtx_jdv_j~e^{-\sum_{j\in\Box_\ell^k}\psi{\left(\frac{\mtx_j}{\ell},v_j\right)}}\Theta\left( \mtx_{j+1}-\mtx_j-a\right),
\label{Z_r-approx-1st}
\end{align}
where $\Box_\ell^k$ represents the phase space corresponding to the $k^{\rm th}$ block and $\{N_k;k=...-1,0,1,...\}$ is the possible number distribution of particles inside these blocks such that $\sum_kN_k=N$. Note that $\{N_k\}$ is a random distribution. {The sum over $j \in \Box^k_\ell$ in the exponent represents sum over rod indices that are within the phase space region $\Box_\ell^k$ {\it i.e.}, over the $N_k$ rods inside  the $k^{\rm th}$ block.} {The approximation in Eq.~\eqref{Z_r-approx-1st}  is based on discarding the contribution from the interaction between the blocks which can be neglected for short-range interaction in the thermodynamic limit.}
Since $ \Psi(\mtx,v)$ given in Eq.~\eqref{Psi(x,v)} is slowly varying, one can assume the function $\Psi(\mtx,v) \approx e^{-\psi(k,v)}$ is more or less uniform in each block and their  values change only across blocks. Using this approximation for large $\ell$ {and introducing the integral representation of the Kronecker delta,} one can write 
\begin{align}
Z_{\rm r}&\approx \int d\mu ~e^{\mu N}\sum_{\{N_k\}}~{e^{-\mu N_k}}\prod_k Z^k_{\rm r}(N_k,\ell),\label{Z_r} \\
\text{with}~~~Z^k_{\rm r}(N_k,\ell)&=\int_{\Box_\ell^k} \prod_jd\mtx_jdv_j~e^{-\sum_{j\in\Box_\ell^k}\psi(k,v_j)}\Theta\left( \mtx_{j+1}-\mtx_j-a\right),\label{def:Z^k-1st}
\end{align}
{where the integral over $\mu$ is along the imaginary axis. In the thermodynamic limit, as shown later, the integral over $\mu$ and the integrals over the configurations will be converted to a saddle-point calculation in which the parameter $\mu$ will serve as a Lagrange multiplier in an appropriate action functional to ensure normalization of the mass.}

Note that $Z^k_{\rm r}(N_k,\ell)$ {in Eq.~\eqref{def:Z^k-1st}} represents the partition function of a homogeneous gas of hard rods in a box of size $\ell$ with particle density $\varrho_k=N_k/\ell$. One can  anticipate that {the partition function $Z^k_{\rm r}(N_k,\ell)$ has a large deviation form:} $Z^k_{\rm r}(N_k,\ell) \asymp e^{-\ell \mt{F}_k(N_k/\ell,\underline{\beta}^k)}$ for large $\ell$ where $\mt{F}_k(\varrho_k,\underline{\beta}^k)$ is the free energy density which is a function of $\varrho_k$ and the set of chemical potentials $\underline{\beta}^k=\{\beta^k_1,\beta^k_2,...\}$. {This large deviation form (represented by  $\asymp$) essentially means the following definition of   the free energy 
\begin{align} 
\mt{F}_k(\varrho_k,\underline{\beta}^k) = - \lim_{ N_k \to \infty \atop \ell \to \infty}~\frac{\ln Z^k_{\rm r}(N_k,\ell) }{\ell}, \notag 
\end{align}
as often found in statistical mechanics \cite{touchette2009large, derrida2007non}.}
Since the velocities are independent, one can again divide the velocity space in the $k^{\rm th}$ block into regions of size $\Delta v$ and  define a velocity distribution function $\mff_{v|k}$ such that $\sum_v \Delta v ~\mff_{v|k}=1$. The partition function of the $k^{\rm th}$ block can now be written as 
\begin{align}
Z^k_{\rm r}(N_k,\ell)&\approx \int...\int d\mff_{v|k}~\exp\left\{- N_k\sum_v \Delta v \left[(\psi(k,v)+\mu)\mff_{v|k} + \mff_{v|k}\ln\left(\frac{N_k\mff_{v|k}}{\ell-aN_k} \right)\right]\right \}, \label{Z^k_r-0}
\end{align}
where the second term in the exponent represents the entropy [see Eq.~\eqref{derv:S_k-a} of ~\ref{ent-hr} for the derivation] corresponding to the macrostate represented by the distribution $\{\mff_{v|k}\}$ . Since $N_k$ is typically large for large $\ell$, one can replace the sums over $v$ as integrals over smooth distribution functions $\mff(v|k)$. We get 
\begin{align}
Z^k_{\rm r}(N_k,\ell)&\approx \int_k \mc{D}[\mff(v|k)]~\exp\left\{-N_k\int dv~ \mff(v|k) \left[ \psi(k,v)+\mu + \ln\left(\frac{N_k\mff(v|k)}{\ell-aN_k} \right)\right] \right \}, \label{Z^k_r}
\end{align}
where $\int_k \mc{D}[\mff(v|k)]$ represents integrals over the velocity distribution profiles inside the $k^{\rm th}$ block (represented by the subscript $k$ of the integration sign). Inserting the expression of $Z^k_{\rm r}(N_k,\ell)$ in Eq.~\eqref{Z_r}, we have 
\begin{align}
Z_{\rm r}&\approx \int d\mu ~e^{\mu N}\sum_{\{N_k\}} \prod_k\int_k \mc{D}[\mff(v|k)]~\exp\left\{- \ell \sum_k \int dv~\mff(k,v) \left[ \psi(k,v)+\mu+ \ln\left(\frac{\mff(k,v)}{1-a\varrho_k} \right)\right]\right \},
\label{Z_r-fn-0}
\end{align}
where ${\mff(k,v)=\mff(v|k)\varrho_k}$.
Approximating the sum over $k$ as integrals over a continuous variable denoted by $y$, we have 
\begin{align}
Z_{\rm r}&\approx \int d\mu ~e^{\mu N}\int \mc{D}[f(y,v)]~\exp\left\{- \ell \int dy \int dv~\mff(y,v) \left[ \psi(y,v)+\mu+ \ln\left(\frac{\mff(y,v)}{1-a\varrho(y)} \right)\right]\right \}, \label{Z_r-fn}
\end{align}
where we identify the sum over the set $\{N_k\}$ and the path integrals in each cell as the path integrals over the phase-space density (PSD) function $\mff(y,v)$ denoted by the measure $\mc{D}[\mff(y,v)]$. Note that the coordinate variable $y$ is measured with respect to the length scale $\ell$ and is related to microscopic coordinate $\mt{y}$ as $\mt{y}=y \ell$. The expression of the partition function $Z_{\rm r}$ in Eq.~\eqref{Z_r-fn} suggests us to identify the following large deviation form for the probability distribution of the PSD $\mff(y,v)$
\begin{align}
\begin{split}
\mc{P}_{\rm r}[\mff(y,v)] &\asymp \frac{1}{Z_r}\int d\mu ~e^{\mu N}~e^{- \ell \mc{F}^{\rm r}_\mu[\mff(y,v)]}= e^{-\ell \mc{F}^{\rm r}_0[\mff(y,v)]}~\delta\left( \ell\int dy \int dv~ \mff(y,v) - N\right),\cr 
\text{where,}~~\mc{F}^{\rm r}_\mu[\mff(y,v)]&= \int dy \int dv~\mff(y,v) \left[ \psi(y,v)+\mu+ \ln\left(\frac{\mff(y,v)}{1-a\varrho(y)} \right)\right].
\end{split}
\label{def:mcP[f]}
\end{align}
It is easy to see that for large $\ell$ the path integral in the expression of the partition function $Z_{\rm r}$ in Eq.~\eqref{Z_r-fn} is given by the saddle-point approximation $Z_{\rm r} \asymp e^{-\ell \mc{F}^{\rm r}_0[\mff_{\rm eq}(y,v)]}$ where $\mff_{\rm eq}(y,v)$ minimizes the free energy $\mc{F}^{\rm r}_0[\mff(y,v)]$ satisfying $\ell \int dy \int dv~\mff_{\rm eq}(y,v)=N$. {The distribution in Eq.~\eqref{def:mcP[f]} describes the probability density of the PSD that varies over scale $\ell$ at time $t=0$ and it  has} the following large deviation form 
\begin{align}
{\mc{P}_{\rm r}[\mff(y,v,0)]} \asymp e^{- \ell \left(\mc{F}^{\rm r}_0[\mff(y,v)]-\mc{F}^{\rm r}_0[\mff_{\rm eq}(y,v)] \right)}~\delta\left( \ell\int dy \int dv~ \mff(y,v) - N\right),
\label{eq:mcP-LDform}
\end{align}
where the $\delta-$ function ensures normalization.

We now study the distribution of an additive observable of the form 
\begin{align}
\mc{B}(\{\mtx_i,v_i\})=\sum_{i=1}^N b(\mtx_i,v_i), \label{add-obs}
\end{align}
in the above statistical state in Eq.~\eqref{eq:mcP-LDform}. The generating function $\msc{Z}^{\mc{B}}_\lambda$, can be defined as 
\begin{align}
\msc{Z}^{\mc{B}}_\lambda = \left \langle e^{-\lambda \mc{B}}\right \rangle_{\mb{P}_{\rm r}}=\prod_{i=1}^N\int d\mtx_i \int dv_i~e^{-\lambda \mc{B}(\{\mtx_i,v_i\})}~\mb{P}_{\rm r}(\{\mtx_i,v_i\}), \label{def:mscZ^A_lam(0)}
\end{align}
where the joint distribution $\mb{P}_{\rm r}$ is given in Eq.~\eqref{eq:mbP_r}. 
Following the same procedure as done for the partition function in Eq.~\eqref{Z_r-fn}, one can also approximate  $\msc{Z}^{\mc{B}}_\lambda $ as path integral over phase-space density profiles. Once again dividing the space into blocks of size $\sim \ell$ and the velocity space into small ranges of size $\Delta v$, and keeping the short-range nature of the interaction in mind, one, in the $N \to \infty$ limit can approximate the multiple integral in Eq.~\eqref{def:mscZ^A_lam(0)} as 
\begin{align}
\msc{Z}_\lambda^{\mc{B}} &\approx  \frac{1}{Z_r} \sum_{\{N_{k}\}} \delta_{\sum_kN_k,N}\prod_k  \int_{\Box^k_\ell} \prod_jd\mtx_j dv_j e^{-\sum_{j\in \Box^k_\ell}( \psi(k,v_j)+\lambda b(\mtx_j,v_j))}   \prod_j\Theta(\mtx_{j+1}-\mtx_j-a) \label{mscZ_lam^B(000)}\\
\begin{split}
& \approx \frac{1}{Z_r}\int d\mu~e^{\mu N} \sum_{{\{N_{k}}\}} \prod_k \int d\mf{f}_{v|k}~ e^{-N_k\sum_v\Delta v \mff_{v|k}\psi(k,v)- \mu N_k}~ \cr 
&~~~~~~~~~~~~~~~~~\times~~e^{-\lambda \ell \sum_v\Delta v \mf{f}_{v|k}\varrho_k\mf{b}(k,v)} \int_{\Box^k_\ell} \prod_jd\mtx_j dv_j \prod_j \Theta(\mtx_{j+1}-\mtx_j-a),
\end{split}
\label{mscZ_lam^B(00)} 
\end{align}
where
\begin{align}
\mf{b}(k,v)= \frac{1}{\ell}\int_{\mt{y}_k-\ell/2}^{\mt{y}_k+\ell/2}d\mt{x}~b(\mt{x},v).
\label{def:mf(b)}
\end{align}
To arrive at Eq.~\eqref{mscZ_lam^B(00)} we have made the following approximations. First we note that the sum $\sum_{j\in \Box_\ell^k}b(\mt{x}_j,v_j)$ in the exponent of Eq.~\eqref{mscZ_lam^B(000)} can be formally written as 
\begin{align}
\sum_{j\in \Box_\ell^k}b(\mt{x}_j,v_j) = 
\sum_v\Delta v \int_{\mt{y}_k-\ell/2}^{\mt{y}_k+\ell/2}d\mt{x}~
~\mt{f}_{\rm r}(\mt{x},v)~b(\mt{x},v),\label{def:sum_b(x,v)}
\end{align}
where {$\mt{f}_{\rm r}(\mtx,u,\mtt) = \sum_{i=1}^N\delta(\mtx -\mtx_i)\delta(v-v_i),$}
is the PSD in microscopic coordinates such that 
{
\begin{align}
 \sum_v~\int_{v-\frac{\Delta v}{2}}^{v+\frac{\Delta v}{2}}du\int_{\mt{y}_k-\frac{\ell}{2}}^{\mt{y}_k+\frac{\ell}{2}}d\mt{x}~\mt{f}_{\rm r}(\mt{x},u)=N_k. \label{eq:sum_v-sum_k-mtf_r}   
\end{align} 
}
\noindent
As mentioned earlier, for the choice of slowly varying statistical state described by Eqs.~\eqref{eq:mbP_r} and \eqref{Psi(x,v)}, the PSD $\mt{f}_{\rm r}(\mt{x},v)$ inside the $k^{\rm th}$ block typically takes the value $\mt{f}_{\rm r}(\mt{x},v) \approx \varrho_k \mf{f}(v|k)$ in the thermodynamic limit {\it i.e.}, in  the limit of large $\ell$ and $N_k$ with $N_k/\ell=\varrho_k$ fixed. Since we are interested in PSD fluctuations on scale $\ell$, we have neglected contributions from fluctuations on shorter length scales. Hence, the sum in Eq.~\eqref{def:sum_b(x,v)} can be approximated as 
\begin{align}
\frac{1}{\ell}\sum_{j\in \Box_\ell^k}b(\mtx_j,v_j) \approx  \sum_v \Delta v~\mf{f}(v|k)\varrho_k\mf{b}(k,v),\label{aprx:sum_b(x,v)}
\end{align}
where $\mf{b}(k,v)$ is defined in Eq.~\eqref{def:mf(b)}.
Note the remaining integrals in Eq.~\eqref{mscZ_lam^B(00)} over microscopic coordinates $\{\mtx_j,v_j\}$ on the product of theta functions essentially provides the entropy factor as in Eq.~\eqref{Z^k_r-0}. We finally get 
\begin{align}
\msc{Z}_\lambda^{\mc{B}} & \approx
 \frac{1}{Z_r}\int d\mu~e^{\mu N} \sum_{\{N_{v|k}\}} \prod_k \int d\mf{f}_{v|k}~ e^{-\ell\sum_k\sum_v\Delta v \mff_{v|k}\varrho_k \left[\psi(k,v)+ \mu  + \lambda \mf{b}(k,v) +\ln\left(\frac{\mff_{v|k}\varrho_k}{1-a\varrho_k} \right)\right]}. 
\label{mscZ_lam^B(0)} 
\end{align}
Taking the appropriate continuum approximations of the sums $\sum_v$, $\sum_k$ and $\sum_{\{N_{v|k}\}}$ [as done to arrive Eq.~\eqref{Z_r-fn}], 
we rewrite $\msc{Z}^{\mc{B}}_\lambda$ in Eq.~\eqref{mscZ_lam^B(0)} as 

\begin{align}
\msc{Z}_\lambda^{\mc{B}} 
& \approx \int \mc{D}[\mff(y,u)]~e^{-\ell \lambda \int dy\int du~\mff(y,u)~\mf{b}(y,u)}~\mc{P}_{\rm r}[{\mff(y,u,0)}],
\label{mscZ_lam^B(0)-1} 
\end{align}
where $\mc{P}_{\rm r}[\mff(0)]$ is given in Eq.~\eqref{eq:mcP-LDform}. This expression suggests us to identify the following representation of observable $\mc{B}$
\begin{align}
\mc{B} &\approx \ell \int dy \int dv~\mf{b}(y,v)~\mff(y,v), \label{redef:mcA} 
\end{align}
as long as we are interested only in the fluctuations at a length scale of order $\sim \ell$ in microscopic units. The above equation further suggests us to define a  density of the fluctuating observables at a local (on the scale of $O(\ell)$) level such as $\mc{B} \approx \int d\mt{y}~\mt{b}(\mt{y})$. Recalling $d \mt{y}=\ell dy$, we can write 
\begin{align}
\mt{b}(\mt{y}=y\ell) \approx \bar{\mf{b}}(y):=\int du~\mf{b}(y,u)~\mff(y,u), \label{def:cg-mta}
\end{align}
from Eq.~\eqref{redef:mcA}.
Once again, we would like to emphasise that the definitions in Eqs.~\eqref{redef:mcA} and \eqref{def:cg-mta} are valid as long as one is interested in the fluctuations  at length scale of $O(\ell)$ in microscopic units. In summary, we find that, for initial configurations chosen from a slowly varying distribution $\mb{P}_{\rm r}$ in Eq.~\eqref{eq:mbP_r}, the statistical state of the system can be described on a length scale of $O(\ell)$ by a slowly varying fluctuating PSD field $\mff(x,v)$ chosen with probability density $\mc{P}_{\rm r}[\mff(0)]$ in Eq.~\eqref{eq:mcP-LDform} and the observables can be expressed in terms of $\mff(x,v)$ as in Eqs.~\eqref{redef:mcA} and \eqref{def:cg-mta}. Now the question is: how does such a statistical state evolve as the rods move microscopically?

As the rods move ballistically, the PSD $\mff(x,v)$ evolve. Since, the the PSD is slowly varying over space, by continuity it also evolves appreciably only over a time scale of $O(\ell)$.
The initial fluctuation described by PSD $\mff(x,v)$, which is different from the mean initial PSD $\mff_{\rm eq}(x,v)$, gets evolved deterministically and produces a fluctuating phase-space density $\mff(x,v,t)$ at later time $\mtt =\ell t$ in microscopic units. In addition to this long wave-length fluctuation, there could be fluctuations due to the non-conserved degrees of freedom. Since the fluctuations in non-conserved degrees of freedom relax quickly \cite{doyon2023ballistic}, the only source of fluctuations that remains in the Euler scaling limit ({\it i.e.}, $\mtx =\ell x,~\mtt=\ell t$) are those which are carried {from the initial state} coherently by the hydrodynamic modes. These modes evolve according to the Euler GHD equation
\begin{align}
\partial_t \mff(x,v,t)+\partial_x\big(v_{\rm eff}[\mff]\mff(x,v,t)\big)=0, \label{eq:eghd}
\end{align}
where 
\begin{align}
v_{\rm eff}[\mff](x,v,t)= \frac{v-a\varrho(x,t)\vartheta(x,t)}{1-a\varrho(x,t)}, \label{def:v_eff}
\end{align}
with $\varrho(x,t)=\int dv ~\mff(x,v,t)$ being the mass density 
and $\vartheta(x,t)$ being the flow velocity of the hard rods  defined by
\begin{align}
\vartheta(x,t)=\frac{\int dv~v\mff(x,v,t)}{\rho(x,t)}. \label{eq:flow-vel-a}
\end{align}
Hence in the Euler scaling limit, the probability distribution of the time {evolution history of the PSD profile $\mff(x,v,t)$ }
is given by 
\begin{align}
\mc{P}_{\rm r}\left[\mff(x,v,t)\right] \asymp \int d\mu~e^{\mu N}e^{- \ell \Delta \mc{F}^{\rm r}_\mu[\mff(x,v,0)] }~\delta\bigg[\partial_t \mff(x,v,t)+\partial_x{\{}v_{\rm eff}[\mff]\mff(x,v,t){\}}\bigg], \label{mcP[mff(t)]}
\end{align}
{for $-\infty < x<\infty$, $-\infty <v<\infty$ and $0<t<\infty$,} where $\Delta \mc{F}^{\rm r}_\mu[\mff]= \mc{F}^{\rm r}_\mu[\mff]-\mc{F}^{\rm r}_\mu[\mff_{\rm eq}]$ and the $\delta[...]$ in the above equation is a $\delta-$functional. The  large deviation form in Eq.~\eqref{mcP[mff(t)]} for the probability distribution of fluctuating field $\mff(x,v,t)$ in the Euler scaling limit provides first main ingredient of the BMFT. The second ingredient comes from the principle of local relaxation \cite{doyon2023emergence, doyon2023ballistic}, according to which the fluctuations of any observables are controlled only by the fluctuations of the conserved densities via hydrodynamic projections. 
More clearly, in order to study the evolution of the fluctuations of an observable like $\mc{B}$ (see Eq.~\eqref{add-obs}) in the Euler scaling limit, one can construct a similar hydrodynamic projection representation as in Eq.~\eqref{redef:mcA} \cite{doyon2023ballistic}
\begin{align}
\mc{B}(\mtt=\ell t) \approx \ell \int dy \int du~\mf{b}(y,u)~\mff(y,u,t). \label{redef:mcA(t)}
\end{align}
Further physical arguments justifying above hydrodynamic representation of additive observable can be found in \cite{doyon2023ballistic}.
From Eq.~\eqref{redef:mcA(t)}, one can define the local observable as 
\begin{align}
\mt{b}(\mt{y}=\ell y, \mtt=\ell t) \approx \bmf{b}(y,t):=\int du~\mf{b}(y,u)~\mff(y,u,t). \label{def:cg-mta(t)}
\end{align}
where note that $\mf{b}(y,u)$ is the same function defined in Eq.~\eqref{def:mf(b)}.
For example, the conserved quantities in Eq.~\eqref{cons-Q} can be written as 
\begin{align}
Q_\nu (\mtt=\ell t)\approx \ell \int dx \int dv ~v^\nu~\mff(x,v,t), \label{def:Q_alpha}
\end{align}
which allows one to identify the density of the conserved quantity $Q_\nu$ as 
\begin{align}
Q_\nu(\mtt=\ell t) \approx \ell \int dx~\mf{q}_\nu(x,t),~~\text{where}~~\mf{q}_\nu(x,t)= \int dv~ v^\nu~\mff(x,v,t). \label{def:q_alpha}
\end{align}
Given the probability distribution functional $\mc{P}_{\rm r}[\mff(t)]$ in Eq.~\eqref{mcP[mff(t)]} and the hydrodynamic representation of  the observable $\mc{B}(\ell t)$ in Eq.~\eqref{def:cg-mta(t)}, the generating function (GF) of $\mc{B}$ can be computed as
\begin{align}
\msc{Z}_\lambda^{\mc{B}}(\mt{T}=\ell T) &= \llangle[\big]e^{-\lambda \mc{B}(\ell T)} \rrangle[\big]_{\mc{P}_{\rm r}}\cr 
&= \int d\mu~ e^{\mu N}\int \mc{D}[\mff(x,v,t)]~e^{-\ell \Delta \mc{F}_{\mu}^{\rm r}[{\mff(x,v,0)}]} e^{-\ell \lambda \int dx\int {du \mf{b}(x,u)\mff(x,u,T)}} \cr
&~~~~~~~~~~~~~~~~~~~~~~~\times ~\delta\bigg[\partial_t \mff(x,v,t)+\partial_xv_{\rm eff}[\mff]\mff(x,v,t)\bigg].~~
\end{align}
The notation $\llangle .... \rrangle_{\mc{P}_{\rm r}}$ represents average over functions (or functionals) of $\mff(x,v,t)$ with respect to the distribution $\mc{P}_{\rm r}[\mff(t)]$. 
Introducing an integral representation of the delta functional through an auxiliary field $\mf{h}(x,v,t)$, one gets 
\begin{align}
&\msc{Z}_\lambda^{\mc{B}}(\mt{T}=\ell T) = \llangle[\big]e^{-\lambda \mc{B}(\ell T)} \rrangle[\big]_{\mc{P}_{\rm r}}= \int d\mu~ e^{\mu N}\int \mc{D}[\mff(x,v,t),\mf{h}(x,v,t)]~e^{-\ell \mc{S}^{\mc{B}}_\lambda[\mff(x,v,t), \mf{h}(x,v,t)]}\cr 
&\text{where,} \label{def:Z_lam} \\
&\mc{S}^{\mc{B}}_{\lambda,\mu}[\mff,\mf{h}]=\Delta \mc{F}^{\rm r}_\mu[\mff(0)]+\lambda \int_0^{\tilde{T}}dt\int dx\int dv \left\{ \mf{b}(x,v)\delta(t-T)\mff(t)+\mf{h}(t)\left[ \partial_t \mff(t)+\partial_xv_{\rm eff}\mff(t)\right] \right\}, \notag
\end{align}
and $\tilde{T}$ is some time larger that $T$. In the above, we have used the short-hand notation $\mff(t)=\mff(x,v,t)$ and $\mf{h}(t)=\mf{h}(x,v,t)$, {and $\Delta \mc{F}^{\rm r}_\mu[\mff]= \mc{F}^{\rm r}_\mu[\mff]-\mc{F}^{\rm r}_\mu[\mff_{\rm eq}]$ with $\mc{F}^{\rm r}_\mu[\mff]$ is given in Eq.~\eqref{def:mcP[f]}}. For large $N$ and $\ell$, the path integral {and the $\mu$-integral can be computed by the saddle-point method. Recall, the $\mu$-integral should be performed as a contour integral on the complex plane. Since, in the limit of large $\ell$, the integral is dominated by the contribution from the saddle-point, we implicitly assume that the contour can be appropriately deformed to pass through the saddle-point \cite{wong2001asymptotic}. Hence, we consider $\mu$ appearing in the action $\mc{S}^{\mc{B}}_{\lambda,\mu}[\mff,\mf{h}]$ through Eq.~\eqref{def:mcP[f]}, as a Lagrange multiplier. We finally get }
\begin{align}
\msc{Z}_\lambda^{\mc{B}}(\ell T) \asymp \exp\left(-\ell~ \mc{F}_{\mc{B}}(\lambda,T)\right),~~\text{with,}~~\mc{F}_{\mc{B}}(\lambda, T)=\mc{S}^{\mc{B}}_{\lambda, {\mu^{\rm sd}}}[\mff^{\rm sd}_\lambda(t),\mf{h}^{\rm sd}_\lambda(t)],
\label{eq:mcF}
\end{align}
where the saddle-point solutions $\mff^{\rm sd}_\lambda(t),~\mf{h}^{\rm sd}_\lambda(t)$ and $ {\mu^{\rm sd}}$ satisfy 
\begin{align}
\left(\frac{\delta \mc{S}^{\mc{B}}_{\lambda, {\mu^{\rm sd}}}}{\delta \mff}\right)_{\mff^{\rm sd}_\lambda,\mf{h}^{\rm sd}_\lambda}=0,
~\left( \frac{\delta \mc{S}^{\mc{B}}_{\lambda, {\mu^{\rm sd}}}}{\delta \mf{h}}\right)_{\mff^{\rm sd}_\lambda,\mf{h}^{\rm sd}_\lambda}=0,
~~\text{and}~~\left( \frac{\partial\mc{S}^{\mc{B}}_{\lambda,\mu}[\mff^{\rm sd}_\lambda,\mf{h}^{\rm sd}_\lambda]}{\partial \mu}\right)_{\mu =  {\mu^{\rm sd}}} =0.
\end{align}
Explicitly, the saddle-point equations are given by 
\begin{subequations}
\label{eq:saddle-a}
\begin{align}
&\partial_t \mff^{\rm sd}_\lambda(x,v,t)+\partial_{x} \big(v_{\rm eff}~ \mff^{\rm sd}_\lambda(x,v,t) \big)=0 \label{eq:sad-2-a}\\
&\partial_t \mf{h}^{\rm sd}_\lambda(x,v,t)+v_{\rm eff}(x,t)\partial_{x} \mf{h}^{\rm sd}_\lambda(x,v,t) \cr 
&~~~~~~~~+\int du\frac{a(v_{\rm eff}-v)}{1-a\varrho(x,t)}\mff^{\rm sd}_\lambda(x,u,t)\partial_{x}\mf{h}^{\rm sd}_\lambda(x,u,t)
=\lambda \mf{b}(x,v)\delta(t-T), \label{eq:sad-1-a}\\
&{\mf{h}^{\rm sd}_\lambda(x,v,\tilde{T})=0,} \label{eq:sad-3-a}\\ 
&\mf{h}^{\rm sd}_\lambda(x,v,0)= \frac{\delta \mc{F}^{\rm r}_\mu[\mff]}{\delta \mff^{\rm sd}_\lambda(x,v,0)}, \label{eq:sad-4-a} 
\end{align}
\end{subequations}
along with the normalization condition. 
\begin{align}
\ell\int dx \int dv~\mff^{\rm sd}_\lambda(x,v,0)=N,  \label{eq:sad-5-a}
\end{align}
and the following boundary conditions;
\begin{align}
\mf{h}^{\rm sd}_\lambda(x,v,t)\big{|}_{x \to \pm \infty}=0,~~\text{and}~~\mathfrak{f}^{\rm sd}_\lambda(x,v,t)\big{|}_{x \to \pm \infty}=0~~\forall~0<t<\tilde{T}. \label{bc-hf-a}
\end{align} 
The second boundary condition is required for the function to normalize as in Eq.~\eqref{eq:sad-5-a}. The normalization condition in Eq.~\eqref{eq:sad-5-a} fixes the chemical constant $ {\mu^{\rm sd}}$. 
By solving the equations \eref{eq:saddle-a} and evaluating $\mc{S}^{\mc B}_{\lambda,\mu}[\mff,\mf{h}]$ on these solutions, one gets the cumulant generating function $\mc{F}_{\mc{B}}(\lambda,T)$ defined in Eq.~\eqref{eq:mcF}. Taking the derivative of $\mc{F}_{\mc{B}}(\lambda,T)$ with respect to $\lambda$, one can compute different cumulants of $\mc{B}(\ell T)$. In this paper, we are {interested in computing the}  two-point-two-time correlation of the PSD using the method described above.

The correlation of PSDs at two macroscopically separated locations at two different times 
is  defined as  
\begin{align}
\mc{C}(\mtx_a,u_a,\mtt_a;\mtx_b,u_b;\mtt_b)&=\langle \mt{f}_{\rm r}(\mtx_a,u_a,\mtt_a)\mt{f}_{\rm r}(\mtx_b,u_b,\mtt_b)\rangle - \langle \mt{f}_{\rm r}(\mtx_a,u_a,\mtt_a) \rangle \langle \mt{f}_{\rm r}(\mtx_b,u_b,\mtt_b)\rangle, 
 \label{def:den-corr}
\end{align}
where, $\mt{f}_{\rm r}(\mtx,u,\mtt)=\sum_{i=1}^N\delta(\mtx-\mtx_i)\delta(v-v_i)$ is the PSD in microscopic coordinates such that $\int d\mtx \int du ~\mt{f}_{\rm r}(\mtx,u,\mtt)=N$. 
The two-point correlation defined in Eq.~\eqref{def:den-corr} can be easily recast as 
\begin{align}
\mc{C}(\mtx_a,u_a,\mtt_a;\mtx_b,u_b;\mtt_b)&= -\left[\frac{\partial}{\partial \lambda} \frac{\langle \mt{f}_{\rm r}(\mtx_b,u_b,\mtt_b)e^{-\lambda \mt{f}_{\rm r}(\mtx_a,u_a,\mtt_a)} \rangle_{\mb{P}_{\rm r}}}{\langle e^{-\lambda {\mt{f}_{\rm r}}(\mtx_a,u_a,\mtt_a)} \rangle_{\mb{P}_{\rm r}}} \right]_{\lambda=0}.
 \label{def:den-corr-1}
\end{align}
For slowly varying initial states described by Eqs.~\eqref{eq:mbP_r} and \eqref{Psi(x,v)}, the above computation of correlation can be converted to a saddle-point calculation  by approximating $\langle ... \rangle_{\mb{P}_{\rm r}}$ with $\llangle ... \rrangle_{\mc{P}_{\rm r}}$. To do so we first notice that 
the PSD $\mt{f}_{\rm r}(\mtx=x\ell,v,\mtt=t\ell)$ typically takes the (mesoscopic) mean value 
\begin{align}
 \mff(x,v,t) \approx \frac{1}{\ell \Delta v} \int_{x\ell-\ell/2}^{x\ell+\ell/2}d\mt{y} \int_{v-\Delta v/2}^{v+\Delta v/2} du~\mt{f}_{\rm r}(\mt{y},u,t\ell),
 \label{approx:mrf_r-mff}
\end{align}
which can be seen from Eq.~\eqref{eq:sum_v-sum_k-mtf_r} by summing both sides over $k$ and taking the continuum limit of the sum to get  $N=\ell \int dy \int dv ~\mff(y,v,t)$. 
As before, the approximation in Eq.~\eqref{approx:mrf_r-mff} captures only the large-scale fluctuations of the PSD over space-time scale of $O(\ell)$ and neglects fluctuations on shorter microscopic scales.  The phase-space density inside a fluid cell centered at $\mtx=x \ell$ can be written as $\mff(x,v,t)=\mff(v|x)\varrho(x,t)$, where $\varrho(x,t) = \int dv~\mff(x,v,t)$ is the mass density inside the fluid cell. Since the mass density varies over the length scale $\ell$,  it is more or less uniform over microscopic coordinates inside the fluid cell. The same approximation has been used earlier in Eq.~\eqref{Z_r-fn-0} to arrive at Eq.~\eqref{Z_r-fn}.

Replacing $\mt{f}_{\rm r}(\mtx=x\ell,v,\mtt=t\ell)$ by its typical value  
$\mff(x,v,t)  =\int dy \int dw~ \delta(x- y)\delta(v-w)~\mff(y,w,t)$
and approximating $\langle ... \rangle_{\mb{P}_{\rm r}}$ by $\llangle ... \rrangle_{\mc{P}_{\rm r}}$ in Eq.~\eqref{def:den-corr-1}, one finds 
\begin{align}
\begin{split}
\mc{C}&(\mtx_a=\ell x_a,u_a,\mtt_a=\ell t_a;\mtx_b=\ell x_b,u_b;\mtt_b=\ell t_b)\cr 
&=-\left[\frac{\partial}{\partial \lambda} \left\{\frac{\int d\mu~ e^{\mu N}\int \mc{D}[\mff(x,v,t),\mf{h}(x,v,t)]~\mff(x_b,u_b,t_b)e^{-\ell \mc{S}^{\mt{f}_{\rm r}}_{\frac{\lambda}{\ell}}[\mff(x,v,t), \mf{h}(x,v,t)]}}
{\int d\mu~ e^{\mu N}\int \mc{D}[\mff(x,v,t),\mf{h}(x,v,t)]~e^{-\ell \mc{S}_{\frac{\lambda}{\ell}}[\mff(x,v,t), \mf{h}(x,v,t)]}}\right\}\right]_{\lambda=0}
\end{split}
 \label{def:den-corr-2}
\end{align}
with
\begin{align}
\mc{S}^{\mt{f}_{\rm r}}_{\lambda}[\mff, \mf{h}]= \Delta \mc{F}^{\rm r}_\mu[\mff(0)]+ \int_0^{\tilde{T}}dt\int dx\int dv \left\{ \lambda\delta(x-x_a)\delta(v-u_a)\delta(t-t_a)\mff(t)+\mf{h}(t)\left[ \partial_t \mff(t)+\partial_xv_{\rm eff}\mff(t)\right] \right\}, \label{def:mcS}
\end{align}
where $\tilde{T}$ is some time larger than $\max(t_a,t_b)$.

For large $\ell$, performing the path integrals using saddle-point method, both in the numerator and the denominator of Eq.~\eqref{def:den-corr-2},  one finds that the correlation $\mc{C}$ has the following scaling form 
\begin{align}
\begin{split}
\mc{C}(\mtx_a=\ell x_a,u_a,\mtt_a=\ell t_a;\mtx_b=\ell x_b,u_b;\mtt_b=\ell t_b)&=\frac{1}{\ell}\msc{C}(x_a,u_a,t_a;x_b,u_b,t_b), \cr
\text{with,}~~~~~~~~\msc{C}(x_a,u_a,t_a;x_b,u_b,t_b)&= -\left[\frac{\partial}{\partial \lambda} \mff_{ \lambda}^{\rm sd}(x_b,u_b,t_b) \right]_{\lambda=0}, 
\end{split}
 \label{def:den-corr-3}
\end{align}
where $\mff_{\lambda}^{\rm sd}(x,u,t)$ for $0<t<\tilde{T}$  is the saddle-point PSD profile that minimises the action $\mc{S}^{\mt{f}_{\rm r}}_{\lambda}$ in Eq.~\eqref{def:mcS}. One now has to solve the saddle-point equations \eqref{eq:saddle-a} with $\mf{b}(x,v)=\delta(x-x_a)\delta(v-u_a)$ with $T=t_a$ which can be done in two ways. One can solve the Eqs.~\eqref{eq:saddle-a} directly in hard-rod picture following a similar method considered in \cite{doyon2023ballistic}. Another way is to formulate an equivalent version of the the BMFT problem using an exact mapping to hard point gas. For completeness and  comprehensiveness, we discuss both ways of solving the BMFT problem. The method using hard-rod picture is presented in~\ref{derv:sad-eq-hr} whereas the solution in the hard point  picture is  presented in the next section.

\section{Computation of the saddle-point solution using a mapping to hard point gas}
\label{BMFT-hp}
To find the saddle-point PSD $\mff^{\rm sd}_\lambda(x,v,t)$, it seems convenient to map the microscopic dynamics of $N$ hard rods to the dynamics of $N$ hard point particles.
For each configuration $\{\mtx_i,v_i\}$ of $N$ hard rods one can construct a configuration $\{\mtx_i',v_i'\}$ of $N$ hard point particles (each of unit mass). The mapping is obtained by removing the non-accessible spaces between successive rods and is given by \cite{lebowitz1968time,percus1969exact}
\begin{align}
\mtx_i'=\mtx_i-(i-1)a,~~v_i'=v_i,~~\text{for}~~i=1,2,...,N. \label{hr-hp-map}
\end{align}
Since for each configuration of the hard-rod gas, there is a unique hard point particle configuration, the ensemble of initial configurations of hard rods can in principle be described by an ensemble of hard point particles and vice-versa. 
The joint distribution $\mathbb{P}_{\rm p}(\{\mtx_i',v_i'\})$ of the initial positions and velocities $\{\mtx_i',v_i'\}$ of the corresponding hard point particles can be easily obtained using this mapping  and one finds
\begin{align}
\mathbb{P}_{\rm p}(\{\mtx_i',v_i'\}) 
&= \frac{1}{Z_{\rm p}}\prod_{i=1}^N\Psi\left(\mtx_i'+a\sum_{j\ne i}\Theta(\mtx'_i-\mtx'_j),v_i\right)\prod_{i=1}^{N-1}\Theta(\mtx'_{i+1}-\mtx'_i), \label{eq:mbP_p}
\end{align}
where, recall, $\Psi(\mtx,v)$ is slowly varying over length scale $\ell$ (see Eq.~\eqref{Psi(x,v)}) and $\Theta(\mtx)$ is the Heaviside theta function. As seen in the case of hard rods in Eq.~\eqref{def:mcP[f]}, for hard point gas also one expects a macroscopic description of the initial state in terms of hard point PSD $\fn(x',v)$ which occurs with probability density
\begin{align}
\mathcal{P}_{\rm p}[\fn(x',v,0)]
&=\int d\mu~e^{\mu N} \exp \left( -\ell \Delta \mathcal{F}^{\rm p}_{\mu}[\fn(x',v',0)] \right),
\label{eq:mcalP} 
\end{align}
where $x'=\frac{\mtx'}{\ell}$ is the macroscopic scale coordinate and $\Delta \mc{F}^{\rm p}_{\mu}=\mc{F}^{\rm p}_{\mu}[\fn]-\mc{F}^{\rm p}_{\mu}[\fn^{\rm eq}]$, with
\begin{align}
\mathcal{F}^{\rm p}_{\mu}[\fn(x',v,0)]&=\int dx' \int dv~\fn(x',v,0)\left\{ \psi(x'+a\Fn(x',0),v) +\mu+\ln \fn(\mtx',v,0)\right\}. \label{eq:fr-enr}
\end{align}
Here $\Fn(x',0)=\int dy' \int dv'~\Theta(x'-y')\fn(y',v',0)$ 
and $\fn^{\rm eq}(x',v)$ is  the equilibrium/stationary  PSD of the hard-point particles that minimises the above functional $\mc{F}^{\rm p}_{\mu}[\fn]$. The last term inside the curly bracket in Eq.~\eqref{eq:fr-enr} appears from the entropy of a gas of free point particles. As these particles move ballistically this statistical state also evolve, however slowly over a time scale of $O(\ell)$. The PSD $\fn(x',v,0)$ evolves according to the Euler equation 
\begin{align}
\partial_t \fn(x',v,t) +v \partial_{x'} \fn(x',v,t)=0. \label{evo-mf}
\end{align}
Given the initial profile $\fn(x',v,0)$, it is easy to solve this equation 
\begin{align}
\fn(x',v,t)=\fn(x'-vt,v,0), \label{sol-mf}
\end{align}
which evidently conserves the total mass $\ell \int dx'\int dv~\fn(x',v,t)=\ell \int dx'\int dv~\fn(x',v,0)= N$.

In the Euler scaling  limit the probability of the history of the PSD $\fn(x',v,t)$ profile over (scaled) time duration $0\leq t \leq \tilde{T}$ is given by 
\begin{align}
\mc{P}_{\rm p}[\fn(x',v,t)]\asymp \int d\mu ~e^{\mu N}~e^{- \ell \Delta \mc{F}^{\rm p}_\mu[\fn(x',v,0)] }~\delta\bigg[\partial_t \fn(x',v,t)+v\partial_{x'}\fn(x',v,t)\bigg],
\label{mcP_p[mff(t)]}
\end{align}
which, after introducing an integral representation of the delta functional, can be rewritten as 
\begin{align}
\begin{split}
\mc{P}_{\rm p}[\fn(x',v,t)] &\asymp \int d\mu ~e^{\mu N} \int \mc{D}[\hn(x',v,t)]~e^{-\ell \msc{S}_\mu[\fn(x',v,t),\hn(x',v,t)]},\cr
\text{where,}~~~\msc{S}_\mu[\fn,\hn] =\Delta \mc{F}^{\rm p}_\mu[\fn(0)] &+\int_0^{\tilde{T}}dt\int dx' \int dv~ \hn(x',v,t)\left\{ \partial_t \fn(x',v,t) +v \partial_{x'} \fn(x',v,t)\right\}. 
\end{split}
\label{def:mscS}
\end{align}
\noindent
The macro-scale coordinate $x'$ and the PSD $\fn(x',v,t)$ of the point particles are related to those of the hard-rod gas via the following transformations \cite{singh2024thermalization, powdel2024conserved}
\begin{align}
\fn(x'_{{\mff(t)}}(x),v,t) = \frac{\mathfrak{f}(x,v,t)}{(1-a \varrho(x,t))},~~\text{with}~~x'_{\mff(t)}(x)=x-a\msc{F}_{\rm r}(x,t), \label{eq:f->mf}
\end{align}
where 
\begin{align}
\varrho(x,t) = \int dv ~\mathfrak{f}(x,v,t),~~\msc{F}_{\rm r}(x,t) = \int^x_{-\infty} dy~\varrho(y,t), \label{def:rho-F}
\end{align}
and 
\begin{align}
\mathfrak{f}(x_{\fn(t)}(x'),v,t) = \frac{\fn(x',v,t)}{(1+a \rho(x',t))},~~\text{with}~~x_{\fn(t)}(x')=x'+a\Fn(x',t). \label{eq:mf->f}
\end{align}
where
\begin{align}
\rho(x',t) = \int dv~ \fn(x',v,t),~~\Fn(x',t) = \int^{x'}_{-\infty} dy'~\rho(y',t). \label{def:rho^0-mf}
\end{align}
{One should carefully note  that $\rho(x',t)$ represents the mass density of the hard point gas and whereas $\varrho(x,t)$ represents the same for the hard-rod gas. See  Table~\ref{tab:variable_comparison} comparing the notations of variables for hard-rod gas and hard point gas.}
\begin{table}[t]
\centering 
    \renewcommand{\arraystretch}{1.75}
\begin{tabular}{|c|c|c|}
    \hline
    Variable & Hard rod & Hard point particle \\ 
    \hline\hline
    Position in & $\mtx$ & $\mtx'$ \\
    microscopic coordinate & & \\
    \hline
    Velocity & $v$ & $v$ \\ 
    \hline
    Time in & $\mtt$ & $\mtt$ \\
    microscopic coordinate & & \\
        \hline
    Position and time in & $x=\frac{\mtx}{\ell},~t=\frac{\mtt}{\ell}$ & $x'=\frac{\mtx'}{\ell},~t=\frac{\mtt}{\ell}$ \\
    macroscopic coordinate & & \\
         \hline
    Mass density & $\varrho(x,t)=\int dv~\mff(x,v,t)$ & $\rho(x',t)=\int dv~\fn(x',v,t)$ \\ 
    \hline
    Cumulative mass density & $\msc{F}_{\rm r}(x,t)=\int_{-\infty}^xdy~\varrho(y,t)$ & $\Fn(x',t)=\int_{-\infty}^{x'}dy'~\rho(y',t)$ \\
    \hline
    Transformation & $x(t) \to x'(t)$ & $x'(t) \to x(t)$ \\
    & $x'_{\mff(t)}(x)=x-a\msc{F}_{\rm r}(x,t)$ & $x_{\fn(t)}(x')=x'+a\Fn(x',t)$\\
    \hline
\end{tabular}
    \caption{{Notations of the variables for hard rods and point particles.}} 
        \label{tab:variable_comparison} 
\end{table}

\noindent
Once PSD $\fn(x',v,t)$ is known from the solution in Eq.~\eqref{sol-mf}, PSD $\mathfrak{f}(x,v,t)$ of hard rods at time $t$ can be obtained  from Eq.~\eqref{eq:mf->f}. 
Recall that the fluctuation of  the initial PSD $\fn(x',v,0)$ or $\mathfrak{f}(x,v,0)$ gives rise to the fluctuation of the phase-space density at time $t$. 
In the Euler scaling limit, one can write  the observable $\mc{B}$ defined in Eq.~\eqref{redef:mcA(t)} in terms of of the point particle PSD $\fn(x',v,t)$  as 
\begin{align}
\mc{B}(\mtt=\ell t) \approx \ell \int dy' \int du~\mf{b}(y'+a\Fn(y',t),u)~\fn(y',u,t), \label{redef:mcA(t)_p}
\end{align}
which allows one to write the GF $\msc{Z}_\lambda^{\mc{B}}$ as path integrals over $\fn(x',v,t)$ and the auxiliary field $\hn(x',v,t)$ : 
\begin{align}
\msc{Z}^{\mc{B}}_\lambda(\ell T) = \llangle[\big] e^{-\lambda \mc{B}(\ell T)}\rrangle[\big]_{\mc{P}_{\rm p}} 
= \int d\mu~ e^{\mu N}\int \mc{D}[\fn(x,v,t),\hn(x,v,t)]~e^{-\ell \mc{S}^{\mc{B}}_\lambda[\fn,\hn]}
\end{align}
where the action now reads 
\begin{align}
\mc{S}_{\lambda}^{\mc{B}} = \msc{S}_\mu[\fn,\hn]  +\lambda \int_0^{\tilde{T}}dt\int dx'\int dv~ \mf{b}(x'+a\Fn(x',t),v)\delta(t-T)\fn(x',v,t),
\end{align}
with $\msc{S}_\mu[\fn,\hn]$ given in Eq.~\eqref{def:mscS} and $\tilde T > T>0$. Similarly, the computation of the two-point-two-time ({\it that is,} unequal space-time) correlation defined in Eq.~\eqref{def:den-corr} can be expressed as path integrals as in Eq.~\eqref{def:den-corr-2} but now over $\fn(x',v,t)$ and $\hn(x',v,t)$ with $\mf{b}(x,v)={\msc{B}(x,v)=}\delta(x-x_a)\delta(v-u_a)$. Once again,  for large $\ell$ the path integral and the integral over $\mu$ can be performed using the saddle-point method. Finally, in the point-particle picture one has to minimise the following action functional
\begin{align}
\mc{S}^{\mt{f}_{\rm r}}_{\lambda}[\fn, \hn]=  \msc{S}_\mu[\fn,\hn]+ \lambda \int_0^{\tilde{T}}dt\int dx'\int dv~\delta(x'+a\Fn(x',t)-x_a)\delta(v-u_a)\delta(t-t_a)\fn(x',v,t).
\label{def:mcS-p}
\end{align}
The saddle-point solutions, denoted by $\fn_\lambda^{\rm sd}(t)$ and $\fn_\lambda^{\rm sd}(t)$ satisfy the following equations 
\begin{subequations}
\label{eq:saddle}
\begin{align}
&\partial_t \fn_\lambda^{\rm sd}(x',v,t)+v\partial_{x'} \fn_\lambda^{\rm sd}(x',v,t)=0 \label{eq:sad-1}\\
&\partial_t \hn_\lambda^{\rm sd}(x',v,t)+v\partial_{x'} \hn_\lambda^{\rm sd}(x',v,t) = \lambda ~\mathcal{R}_{\mff^{\rm sd}_\lambda(t)}[\msc{B}(x_{\fn_\lambda^{\rm sd}(t)}(x'),v)]\delta(t-t_a), \label{eq:sad-2}\\
&\hn_\lambda^{\rm sd}(x',v,\tilde{T})=0 \label{eq:sad-3}\\ 
&\hn_\lambda^{\rm sd}(x',v,0)= \left(\frac{\delta \mathcal{F}^{\rm p}_\mu[\fn(0)]}{\delta \fn(x',v,0)}\right)_{\fn(0)=\fn_\lambda^{\rm sd}(0)}, \label{eq:sad-4} 
\end{align}
\end{subequations}
where
\begin{align}
\msc{B}(x,v)=\delta(x-x_a)\delta(v-u_a). \label{def:mscA}
\end{align}
The derivation of the saddle-point equations \eqref{eq:saddle} are provided in \ref{derv:sad-eq}. We refer the equations \eqref{eq:saddle} as the BMFT equations. 
Recall from  Eq.~\eqref{eq:mf->f} that $x_{\fn(t)}(x')=x'+a\Fn(x',t)$ and $\mff^{\rm sd}_\lambda(x_{\fn_\lambda^{\rm sd}(t)}(x'),v,t)$ is the saddle-point  PSD of hard rods corresponding to the saddle-point  PSD $\fn_\lambda^{\rm sd}(t)$ of point particles. In the above, we have defined the operation $\mathcal{R}_{\mff(t)}$ acting on a function $H(x,v)$ as 
\begin{align}
\begin{split}
\mathcal{R}_{\mff(t)}[H(x,v)]&=H(x,v) +a \int dy\int du~\Theta(y-x)\left\{ \partial_{y} H(y,u)\right\}\mff(y,u,t). 
\end{split}
\label{R[H]}
\end{align}
The saddle-point equations should satisfy the boundary conditions. 
\begin{align}
\fn_\lambda^{\rm sd}(x' \to \pm \infty,v,t)  \to 0,~~\forall~t. \label{bc:sad-eq-1}
\end{align} 
The boundary conditions for $\hn_\lambda(x',v,t)$ depend on the choices of the observable. The functional derivative of $\mathcal{F}^{\rm p}_\mu[\fn(0)]$ can be computed from Eq.~\eqref{eq:fr-enr} and it is given by 
\begin{align}
\begin{split}
\frac{\delta \mathcal{F}^{\rm p}_\mu[\fn(0)]}{\delta \fn(x',v,0)}= \mathcal{R}_{\mff(0)}[\psi(x_{{\fn(0)}}(x'),v)] +\ln \fn(x',v,0)+\mu.
 \end{split}
 \label{delmcalF/delf}
\end{align}
The constant $\mu$ is fixed from the normalization $\ell \int dx' \int dv~\fn_\lambda^{\rm sd}(x',v,0)=N$.
Note that if the initial PSD $\fn_\lambda^{\rm sd}(x',v,0)$ satisfies this normalization condition then, by virtue of the equation (\ref{eq:sad-1}), the PSD at a later time will also satisfy the normalization condition. 

Given ${\fn^{\rm sd}_{\lambda}}(x',v,0)$, one solves the saddle-point equations \eqref{eq:saddle} to find ${\fn^{\rm sd}_{\lambda}}(x,v,t)$. This solution is then transformed into hard rod PSD $\mff^{\rm sd}_{\lambda}(x,v,t)$ using the transformation in Eq.~\eqref{eq:mf->f}. Using the solution $\mff^{\rm sd}_{\lambda}(x,v,t)$  in Eq.~\eqref{def:den-corr-3} one would find the scaled correlation $\msc{C}(x_a,u_a,t_a;x_b,u_b,t_b)$.

\section{Solution of the saddle-point equations }
\label{sec:den-corr}
In this section we solve the saddle-point equations \eqref{eq:saddle} with $\msc{B}(x,v)=\delta(x-x_a)\delta(v-u_a)$.  For notational simplicity, from now on we omit the superscript `sd' from the saddle-point functions.
We assume $\tilde{T}$ is some arbitrary time larger than $t_a$ {\it i.e.}, $0\leq t_a <\tilde{T}$. Integrating both sides of Eq.~\eqref{eq:sad-2} {over time  from $t_a -\epsilon$ to $t_a + \epsilon$ with $\epsilon >0$}, we get the value of $\hn_\lambda(x',v,t_a^-)= {\hn_\lambda(x',v,t_a-\epsilon)|_{\epsilon \to 0}}$ as 
\begin{align}
{\hn_\lambda}(x',v,t_a^-) &= \lambda~\delta \mt{g}_{t_a}(x_{\fn_\lambda(t_a)}(x'),v),\label{h(ta-)}
\end{align}
{because $\hn_\lambda(x',v,t>t_a)=0$ which can be seen by evolving Eq.~\eqref{eq:sad-2} backward in time from $\tilde{T}$ to $t_a$ and noting $\hn_\lambda(x',v,\tilde{T})=0$ from Eq.~\eqref{eq:sad-3}. } 
{In the above  we have defined} 
\begin{align}
 \delta \mt{g}_{t_a}(x,v)&= -\mathcal{R}_{{\bmf{f}(t_a)}}[\msc{B}(x,v)]=-\partial_{x_a}\left[ \Theta(x_a-x)\left\{\delta(v-u_a) - a\bmf{f}(x_a,u_a,t_a)\right\} \right]. \label{def:mtg_t_a}
\end{align}
Now, evolving Eq.~\eqref{eq:sad-2} from $t=t_a^-$ to $t=0$ backwards, we get 
\begin{align}
{\hn_\lambda}(x',v,0)={\hn_\lambda}(x'+vt_a,v,t_a) = \lambda~\delta \mt{g}_0(x',v), ~~\text{with}~~\delta \mt{g}_0(x',v)=\delta \mt{g}_{t_a}(x_{\fn_\lambda(t_a)}(x'+vt_a),v).    \label{h(0)}
\end{align}
Using this on the left hand side (lhs) of Eq.~\eqref{eq:sad-4} and performing the functional derivative of $\mathcal{F}^{\rm p}_\mu[\fn_\lambda(0)]$ in Eq.~\eqref{delmcalF/delf} along with Eq.~\eqref{R[H]}, we get 
\begin{align}
\mathcal{R}_{\mff_{\lambda}(0)}[\psi(x_{\fn_\lambda(0)}(x'),v)]+\ln \fn_\lambda(x',v,0)+\mu=\lambda~\delta \mt{g}_{0}(x',v) =\lambda~\delta \mt{g}_{t_a}(x_{\fn_\lambda(t_a)}(x'+vt_a),v).\label{eq:sad-4-dc-1}
\end{align}
One must solve this equation to find $\fn_\lambda(x',v,0)$, which will now depend on $\lambda$ and the observable $\msc{B}(x,v)$. In simple words, what we have done is the following: To see a particular fluctuation in PSD at $x=x_a$ at $t=t_a$, one needs to know what was the particular initial random PSD profile that will evolve to the desired fluctuation at $t=t_a$ and $x=x_a$. To find that initial random profile, we have to evolve equation \eqref{eq:sad-2} backward to obtain $\hn_\lambda(x',v,0)$. This solution is then inserted into Eq.~\eqref{eq:sad-4}, solving which would give the particular initial (random) PSD profile that under forward evolution would give rise to the desired fluctuation at $x=x_b$ at $t=t_b$. 
The density fluctuations at different locations at $t=t_a$ are now correlated because they originate from the same random initial profile, evolved ballistically [Eq.~\eqref{eq:sad-1}] to the desired time. 

To proceed we need to solve Eq.~\eqref{eq:sad-4-dc-1} to find $\fn_\lambda(x',v,0)$ or equivalently $\mff_\lambda(x,v,0)$ for given $\psi(x,v)$, which can be solved order by order in $\lambda$. For that we first show that Eq.~\eqref{eq:sad-4-dc-1} can be rewritten in a simpler way in terms of the hard rod phase-space density $\mff_\lambda(x,v,0)$ as [see ~\ref{derv:f(x,v,0)} for details]
\begin{align}
\ln \left( \frac{\mathfrak{f}_\lambda(x,v,0)}{1-a\varrho_\lambda(x,0)}\right) = -\psi(x,v) -\frac{a\varrho_\lambda(x,0)}{1-a\varrho_\lambda(x,0)}-\mu + \lambda  \mathscr{R}_{\mff_{\lambda}(0)} \left[\delta \mt{g}_{0}(x'_{\mff_\lambda(0)}(x),v) \right].
\label{eq-to-find-mfrkf(0)}
\end{align}
where we have introduced the operation $\mathscr{R}_{\mff(t)}[H(x,v)]$
 \begin{align}
 \begin{split}
\mathscr{R}_{\mff(t)}[H(x,v)]&=H(x,v) -a \int dy~\Theta(y-x)~\left[ \int du\left\{ \partial_{y} H(y,u)\right\}\frac{\mff(y,u,t)}{1-a\varrho(y,t)}\right].
\end{split}
\label{Rsc[H]}
\end{align}
Note the difference between the operations of $\mathscr{R}$ and $\mathcal{R}$ given in Eqs.~\eqref{Rsc[H]} and \eqref{R[H]}, respectively. In fact, it is straightforward to show that the operator $\msc{R}_{\mff(t)}$ is inverse of the
operator $\mc{R}_{\fn(t)}$ {\it i.e.} $\msc{R}_{\mff(t)}[\mc{R}_{\mff(t)}[H(x,v)]] =H(x,v)$ (see ~\ref{sec:mscRmcR=I}).

Equation \eqref{eq-to-find-mfrkf(0)} can be solved for $\mathfrak{f}_\lambda(x,v,0)$ order by order in $\lambda$. To compute the two-point correlation, it suffices to find the solution to linear order in $\lambda$. 
Writing \begin{align}
\mathfrak{f}_\lambda(x,v,0) \simeq \bar{\mathfrak{f}}(x,v,0)+ \lambda \delta \mathfrak{f}(x.v.0),~~~\text{and}~~\mu=\bar{\mu}+\lambda \delta \mu, \notag 
\end{align}
in Eq.~\eqref{eq-to-find-mfrkf(0)} and expanding both sides in powers of $\lambda$ we get 
\begin{align}
&\text{at}~O(\lambda^0):~
 \bmf{f}(x,v,0) =\bar{\mc{N}}~(1-a\brvr(x,0))~e^{-\psi(x,v)} e^{-\frac{a\brvr(x,0)}{1-a\brvr(x,0)}},~~\text{with}~~~\bar{\mc{N}}=e^{-\bar{\mu}},\label{eq:bmf(0)} \\
&~~\text{and} \notag \\
&\text{at}~O(\lambda):~\delta \mff(x,v,0) +\left[a \frac{2-a\brvr(x,0)}{(1-a\brvr(x,0))^2}\delta \varrho(x,0) +\delta {\mu}\right]\bmf{f}(x,v,0)= \delta \mt{h}_{t_a}(x,v)\bmf{f}(x,v,0),\label{eq:dbmf(0)} \\
\begin{split}
&~~\text{where}~~~\delta \mt{h}_{t_a}(x,v) = \mathscr{R}_{\bmf{f}(0)} \left[\delta \mt{g}_0( x'_{\bmf{f}(0)}(x),v)\right],\\
&~~\text{with}~~~~\delta \mt{g}_0(x'_{\bmf{f}(0)}(x),v)=\delta \mt{g}_{t_a}(x_{\brf(t_a)}(x'_{\bmf{f}(0)}(x)+vt_a),v),
\end{split}
 \label{def:mth-0} 
\end{align}
$\brf(x'_{\bmf{f}(0)}(z),v,0)=\frac{\bmf{f}(z,v,0)}{1-a\brvr(z,0)}$ and  $\delta \mt{g}_{t_a}(y,w)$ given in Eq.~\eqref{def:mtg_t_a}.
The constants $\bar{\mu}$ and $\delta \mu$ should be determined from the conditions 
\begin{align}
&\int dv~\bmf{f}(x,v,0)= \bar{\varrho}(x,0), \label{fix-balpha}
\end{align}
and
\begin{align}
&\int dx \int dv~\delta \mff(x,v,0)=0, \label{fix-dalpha}
\end{align}
respectively, {where $\bar{\varrho}(x,0)$ is the initial mass density profile.}


\subsection{Initial average PSD $\bmf{f}(x,v,0)$:} The solution of Eq.~\eqref{eq:bmf(0)} for $\bmf{f}(x,v,0)$ provides the initial average PSD. However, one needs to find $\brvr(x,0)$ and $\bar{\mc{N}}$. Integrating over $v$ on both sides of Eq.~\eqref{eq:bmf(0)}, one gets 
\begin{align}
\frac{\brvr(x,0)}{1-a\brvr(x,0)} =\bar{\mc{N}}\Phi(x) e^{-\frac{a\brvr(x,0)}{1-a\brvr(x,0)}},~~\text{with}~~\Phi(x)=\int dv~e^{-\psi(x,v)}=\int dv~\Psi(x,v), \label{eq:-vrho(0)}
\end{align}
where we have used Eq.~\eqref{Psi(x,v)}.
The solution of Eq.~\eqref{eq:-vrho(0)} can be written in terms of Lambert's $W$ function \cite{dence2013brief}
\begin{align}
\frac{a\brvr(x,0)}{1-a\brvr(x,0)}  = W\left( \bar{\mc{N}} a \Phi(x)\right),~~\implies ~~\brvr(x,0)= \frac{1}{a} \frac{W\left( \bar{\mc{N}} a \Phi(x)\right)}{1+W\left( \bar{\mc{N}} a \Phi(x)\right)}, 
\label{sol:vrho-Lf}
\end{align}
where the normalization constant $\bar{\mc{N}}$ needs to be fixed from
$\ell \int dx~\brvr(x,0)=N$. 
We get
\begin{align}
\bar{\mc{N}}\int dx~(1-a\brvr(x,0))~\Phi(x)~ e^{-\frac{a\brvr(x,0)}{1-a\brvr(x,0)}}=\bar{\mc{N}} \int dx \frac{\Phi(x)~e^{-W\left( \bar{\mc{N}} a \Phi(x)\right)}}{1+W\left( \bar{\mc{N}} a \Phi(x)\right)}=\frac{N}{\ell}.
\end{align}
solving which one can get $\bar{\mc{N}}$. 
Once $\brvr(x,0)$ and $\bar{\mc{N}}$ are determined, inserting them in Eq.~\eqref{eq:bmf(0)} gives the initial average PSD $\bmf{f}(x,v,0)$ explicitly.

\subsection{The desired initial fluctuation in $\mff(x,v,0)$:}
Equation \eqref{eq:dbmf(0)} can be simplified further as follows. Noting that $\delta \varrho(x,0)= \int dv~\delta \mff(x,v,0)$, one first finds $\delta \varrho(x,0)$ by integrating both sides of Eq.~\eqref{eq:dbmf(0)} and then substituting it back in Eq.~\eqref{eq:dbmf(0)} one gets,
\begin{align}
\delta \mff_{t_a}(x,v,0) =\bmf{f}(x,v,0) \left[\delta \mt{h}_{t_a}(x,v) - a(2-a\brvr(x,0))\int du~\bmf{f}(x,u,0)\delta \mt{h}_{t_a}(x,u)-\delta {\mu}(1-a \brvr(x,0))^2 \right].~ \label{sol:dbmf(0)}
\end{align}
The subscript $t_a$ denotes the dependence of $\delta \mff_{t_a}(x,v,0)$ on the phase space coordinates $x_a,u_a$ at time $t_a$. However, for notational simplicity, we drop this dependence explicitly in the following.
To determine $\delta {\mu}$, we insert the expression of $\delta \mff(x,v,0)$ in Eq.~\eqref{sol:dbmf(0)}. We get 
\begin{align}
\int dx~\left[ \int du~\bmf{f}(x,u,0)\delta \mt{h}_{t_a}(x,u) -\brvr(x,0)~\delta {\mu}\right] (1-a\brvr(x,0))^2=0,
\end{align}
which provides
\begin{align}
\delta \mu =  \frac{\int dx \int du~\bmf{f}(x,u,0)\delta \mt{h}_{t_a}(x,u)}{\int dx~\brvr(x)(1-\brvr(x))^2}.
\end{align}
We observe that in the  $N \to \infty$ limit $\delta \mu \to 0$. Hence, from now on we neglect $\delta \mu$ in our subsequent calculations. Consequently, from Eq.~\eqref{sol:dbmf(0)}, we get
\begin{align}
\mff_\lambda(x,v,0)&= \bmf{f}(x,v,0) + \lambda ~\delta \mff(x,v,0)  + O(\lambda^2), \label{sol:bmf(0)-fn}\\
\text{with}~~~\delta \mff(x,v,0) &=  \bmf{f}(x,v,0) \left[\delta \mt{h}_{t_a}(x,v) - a(2-a\brvr(x,0))~\int du~\bmf{f}(x,u,0)\delta \mt{h}_{t_a}(x,u) \right], \label{sol:dbmf(0)-fn}
\end{align}
where $\delta \mt{h}_{t_a}(x,v)$ and $\bmf{f}(x,v,0)$ are given in Eq.~\eqref{def:mth-0} and Eq.~\eqref{eq:bmf(0)}, respectively. 
As shown in Eq.~\eqref{dmf(f)-a} of~\ref{derv:sad-eq-hr}, the expression of $ \delta \mf{f}(x,v,0)$ in Eq.~\eqref{sol:dbmf(0)-fn} can also be obtained by performing the full computation entirely on the hard-rod picture ( {\it i.e.} without  transforming to the point-particle picture) except for the fact that $\delta \mt{h}_{t_a}(x,v)$ should be replaced by $\delta \mf{h}(x,v,0)= -\hmsc{R}_{\bmf{f}(0)}[\hmc{R}_{\bmf{f}(t_a)}[\msc{B}](x_{{\brf}(t_a)}(x'_{\bmf{f}(0)}(x)+vt_a),v)]$. Note that $ \delta \mf{h}(x,v,0)$ is defined in terms of two operators $\hmsc{R}_{\bmf{f}(0)}$ and $\hmc{R}_{\bmf{f}(t_a)}$ similar to operators $\msc{R}_{\bmf{f}(0)}$ and $\mc{R}_{\bmf{f}(t_a)}$. It turns out that $\delta \mt{h}_{t_a}(x,v)=\delta \mf{h}(x,v,0)$ [see~\ref{find_df_r(0)-hr}]. This is proved in Eq.~\eqref{eq:h_ta=-mfh} of~\ref{h_ta=-mfh}.

\subsection{Computation of $\bmf{f}(x,v,t)$}
\label{sec:evo-bmf(f)}
In order to compute the correlation $\msc{C}$ defined in Eq.~\eqref{def:den-corr-3} we need to evolve the PSD $\mff_\lambda(x,v,0)$ to time $t=t_b$ which can be done in the point-particle picture. For that, one requires to find $\fn_\lambda(x',v,0)$, which can be obtained by using the transformation in Eq.~\eqref{eq:f->mf}: 
\begin{align}
\fn_\lambda(x'_{\mff_\lambda(0)}(x),v,0)=\frac{\mff_\lambda(x,v,0)}{1-a\varrho^\lambda(x,0)}.
\end{align}
This equation is essentially an identity between the two functions $\fn_\lambda(x',v,0)$ and $\mff_\lambda(x,v,0)$: 
\begin{align}
\fn_\lambda(x',v,0)=\frac{\mff_\lambda(x_{\fn_\lambda(0)}(x'),v,0)}{1-a\varrho^\lambda(x_{\fn_\lambda(0)}(x'),0)}. \label{eq:f_r^lam(0)->f^lam(0)}
\end{align}
Now one needs to  evolve $\fn_\lambda(x',v,t)$ according to Eq.~\eqref{eq:sad-1} until time $t$. At time $t$, one then has $\fn_\lambda(x',v,t)$, which then needs to be transformed back to PSD $\mff_\lambda(x,v,t)$ of hard rods at time $t$. Schematically, one can follow the flow diagram below:
\begin{align}
\mff_\lambda(x,v,0) \overset{\text{Eq.~\eqref{eq:f->mf}}}{\Longrightarrow} \fn_\lambda(x'_{\mff_\lambda(0)}(x),v,0) \overset{evolve}{\longrightarrow} \fn_\lambda(x'_{\mff_\lambda(0)}(x),v,t) \overset{\text{Eq.~\eqref{eq:mf->f}}}{\Longrightarrow} \mff_\lambda(x_{\fn_\lambda(t)}(x'_{\mff_\lambda(0)}(x)),v,t), \notag
\end{align}
to obtain $\mff_\lambda(x,v,t)$. In order to do so, we first note that the evolution of PSD of hard point particles can be obtained by solving Eq.~\eqref{eq:sad-1}  which is simply given by 
\begin{align}
\fn_\lambda(x',v,t) = \fn_\lambda(x'-vt,v,0). \label{sol:f(t)} 
\end{align}
Transforming back to hard rod PSD at time $t$ using Eqs.~(\ref{eq:f->mf} - \ref{def:rho^0-mf}), one can write the following identity,
\begin{align}
\mff_\lambda(z,v,t) = \frac{\fn_\lambda(x'_{\mff_\lambda(t)}(z),v,t)}{1+a\rho^\lambda(x'_{\mff_\lambda(t)}(z),t)},~~~\text{with}~~x'_{\mff_\lambda(t)}(z) = z- a \msc{F}_{\rm r}^\lambda(z,t). \label{eq:f_f^lam(t)-->f_r^lam(t)-mt}
\end{align}
Recall that for the computation of the space-time correlation, we need to compute $\mff_\lambda(x,v,t)$ to linear order in $\lambda$ only, {\it i.e.}, 
\begin{align}
\mff_\lambda(z,v,t)=\bmf{f}(z,v,t) ~+~ \lambda~\delta \mff(z,v,t) + O(\lambda^2),~~~\text{where}~~\delta \mff(z,v,t) = \left[ \partial_\lambda \mff_\lambda(z,v,t)\right]_{\lambda=0},
\end{align}
and $\bmf{f}(x_{\bar{\fn}(t)}(z'),v,t)=\frac{\bar{\fn}(z',v,t)}{1+a\brr(z',t)}$. Performing some manipulations (as shown in~\ref{derv:df_r}) we obtain
\begin{align}
\delta \mff(z,v,t) 
&=\bmf{f}(z,v,t)\delta \mt{h}_{t_a}^{\rm dr}(x_{{\brf}(0)}(x'_{\bmf{f}(t)}(z)-vt),v) - a \partial_z \left[ \bmf{f}(z,v,t) \delta F(x'_{\bmf{f}(t)}(z),t)\right] \displaybreak[3] \cr 
&~~~~~~~~~~~~~~~~~~~~~~~~~~
+a \partial_{z} \left[ \frac{\bmf{f}(z,v,t)}{1-a\brvr(z,t)}\right]\delta \msc{F}_{\rm r}(x_{\brf(0)}(x'_{\bmf{f}(t)}(z)-vt),0), \displaybreak[3] \label{eq:dmscf(t)}
\end{align}
where
  \begin{align}
\delta \mt{h}_{t_a}^{\rm dr}(x,v) =& \delta \mt{h}_{t_a}(x,v) - a\int du~\bmf{f}(x,u,0)~\delta \mt{h}_{t_a}(x,u). \label{mth_t_a^dr}
 \end{align}
and 
\begin{align}
\delta F(z',t)&= \int dy~\Theta(x_{\brf(t)}(z')-y)\int du~\bmf{f}(y,u,t)~\mt{h}_{t_a}^{\rm dr}(x_{{\brf}(0)}(x'_{\bmf{f}(t)}(y)-ut),u) \cr 
&~~~~~~~~~~~~~
+a\int dy\Theta(x_{\brf(t)}(z')-y) \int du~\partial_{y} \left( \frac{\bmf{f}(y,u,t)}{1-a\brvr(y,t)}\right)\delta \msc{F}_{\rm r}(x_{\brf(0)}(x'_{\bmf{f}(t)}(y)-ut),0),
 \label{def:dF(t)}
\end{align}
with $\delta \msc{F}_{\rm r}(z,0)=\int dy~ \Theta(z-y)\int du ~\delta \mff(y,u,0)$.
In the next section, we will use the expression of $\delta \mff(z,v,t)$ in Eq.~\eqref{eq:dmscf(t)} to compute the two-point correlation.


\section{Computation of the correlation $\msc{C}$ in Eq.~\eqref{def:den-corr-3}:}
\label{sec:eva-corr}
Inserting $\mff(x_b,u_b,t_b)=\bmf{f}(x_b,u_b,t_b)+\lambda~ \delta \mf{f}(x_b,u_b,t_b)$ in  Eq.~\eqref{def:den-corr-2}, 
we obtain $\msc{C}(x_a,u_a,t_a;x_b,u_b,t_b) = - \delta \mf{f}(x_b,u_b,t_b)$ {\it i.e.},
\begin{align}
\begin{split}
\msc{C}(x_a,&u_a,t_a;x_b,u_b,t_b) 
= \bmf{f}(x_b,u_b,t_b)\delta \mt{h}_{t_a}^{\rm dr}(x_{{\brf}(0)}(x'_{\bmf{f}(t_b)}(x_b)-u_bt_b),u_b)  \cr 
&~~~~~~~
- a \partial_{x_b} \left[ \bmf{f}(x_b,u_b,t_b) \delta F(x'_{\bmf{f}(t_b)}(x_b),t_b)\right]
+a \partial_{x_b} \left[ \frac{\bmf{f}(x_b,u_b,t_b)}{1-a\brvr(x_b,t_b)}\right]\delta \msc{F}_{\rm r}(x_{\brf(0)}(x'_{\bmf{f}(t_b)}(x_b)-u_bt_b,0).
\end{split}
 \label{mcS_a_b}
\end{align}
where we have used the explicit expression of $ \delta \mf{f}(x_b,u_b,t_b)$ from Eq.~\eqref{eq:dmscf(t)}.  Recall from Eq.~\eqref{mth_t_a^dr} that the function $\mt{h}_{t_a}^{\rm dr}(z,v)$ can be defined in terms of the function $\mt{h}_{t_a}(x,v)$. 
From this correlation, one can compute the two-point correlation between any two conserved densities $\mf{q}_\nu(x,t)$ (defined in Eq.~\eqref{def:q_alpha}) as 
\begin{align}
\begin{split}
{\mc{C}_{\nu,\nu'}}(x_a,t_a;x_b,t_b)&=\langle \mf{q}_\nu(x_a,t_a)\mf{q}_{\nu'}(x_b,t_b)\rangle - \langle \mf{q}_\nu(x_a,t_a) \rangle \langle \mf{q}_{\nu'}(x_b,t_b)\rangle \approx \frac{1}{\ell} \msc{C}_{\nu,\nu'}(x_a,t_a;x_b,t_b)\\
\text{where,}~~~ \msc{C}_{\nu,\nu'}&(x_a,t_a;x_b,t_b)= \int du_a \int du_b~u_a^\nu u_b^{\nu'}~\msc{C}(\mtx_a,u_a,\mtt_a;\mtx_b,u_b;\mtt_b).
\end{split}
\label{def:qq-corr}
\end{align}
One can also use the correlation $\msc{C}(x_a,u_a,t_a;x_b,u_b,t_b)$ to compute correlations between local observables $\bmf{b}(y,t)$ defined in Eq.~\eqref{def:cg-mta(t)} as 
\begin{align}
\llangle \bmf{b}_1(x,t_a) \bmf{b}_2(y,t_b) \rrangle_{\mc{P}_{\rm r},c} = \int du \int dv ~\mf{b}_1(x,u) \mf{b}_2(y,v)~\msc{C}(x,u,t_a;y,v,t_b),
\end{align}
where subscript `c' denotes connected correlation.
The expression of the correlation $\msc{C}$ in Eq.~\eqref{mcS_a_b} becomes simpler for different special cases, such as for the initial correlation, the equilibrium space-time correlation, and the unequal space-time mass density correlation. We discuss these cases separately in the following.

\subsection{Initial two-point correlation $\msc{C}(x_a,u_a,0;x_b,u_b;0)$: }
For $t_a=t_b=0$ one can easily see that $\delta \mt{h}_0(x,v) =-\delta(v-u_a)\delta(x-x_a)$. This simplifies the expression of $\msc{C}$ in Eq.~\eqref{mcS_a_b} and one has 
\begin{align}
\msc{C}(x_a,u_a,0;x_b,u_b,0) 
&=\delta(x_b-x_a)\bmf{f}(x_b,u_b,0) \left[\delta(u_b-u_a)- a(2-a\brvr(x_b,0))~\bmf{f}(x_a,u_a,0)  \right].
\label{mcS(0,0)}
\end{align}
Since the initial state by assumption is uncorrelated, one does not have any long-range correlation at $t_a=t_b=0$. Also, the above expression is consistent with the expression of two-point correlation given in Eq.~(38) of \cite{doyon2017dynamics}.
Inserting this correlation further into Eq.~\eqref{def:qq-corr} and performing the integrals over $u_a$ and $u_b$, one gets the initial correlation of the conserved densities
\begin{align}
\begin{split}
{\msc{C}_{\nu,\nu'}(x_a,{0};x_b,{0})}&=\delta(x_b-x_a)\left[ \msc{U}_{\nu+\nu'}(x_a,0) -a(2-a\brvr(x_b,0) \msc{U}_{\nu}(x_a,0) \msc{U}_{\nu'}(x_b,0)\right],\\ 
\text{where,}~~~~~~&~~~\msc{U}_{\nu}(x,0) = \int du ~u^\nu~\bmf{f}(x,u,0).
\end{split}
\label{def:qq-corr-ini}
\end{align}
Explicit expressions can be obtained after performing the integral over $u$ in the definition of $\msc{U}_\nu$. For example, the initial correlation between mass densities at two-points $x_a$ and $x_b$ can be obtained explicitly
\begin{align}
\msc{C}_{0,0}(x_a,0;x_b;0)=\brvr(x_b,0)(1-a\brvr(x_b,0))^2\delta(x_a-x_b), \label{fex:s_00(0)}
\end{align}
which reproduces the result derived in \cite{doyon2023ballistic}.

\subsection{Un-equal spacetime correlation $\msc{C}(x_a,u_a,t_a;x_b,u_b;0)$: }
\label{sec:unequal-C}
For $t_b=0$, the correlation is given by $\msc{C}(x_a,u_a,t_a;x_b,u_b;0)= - \delta \mf{f}(x_b,u_b,0)$ which according to Eq.~\eqref{sol:dbmf(0)-fn} is given by 
\begin{align}
\msc{C}(x_a,u_a,t_a;x_b,u_b,0)= -  \bmf{f}(x_b,u_b,0) \left[\delta \mt{h}_{t_a}(x_b,u_b) - a(2-a\brvr(x_b,0))~\int du~\bmf{f}(x_b,u,0)\delta \mt{h}_{t_a}(x_b,u) \right], \label{ex:corr-ue-t-1}
\end{align}
where the integral in the above equation is given in Eq.~\eqref{int-f_r(w)h^dr(w)}. Hence, the mass density correlation is given by 
\begin{align}
\msc{C}_{0,0}(x_a,t+a;x_b,0)=-(1-a \brvr(x_b,0))^2~\int du_a~\int du~\bmf{f}(x_b,u,0)\delta \mt{h}_{t_a}(x_b,u). 
\end{align}
Note that for $t_a=0$ the expression {above} indeed reduces to the initial space correlation in Eq.~\eqref{fex:s_00(0)}.

\paragraph{In equilibrium:} In case of a homogeneous equilibrium state, the expression of the two-point correlation in Eq.~\eqref{ex:corr-ue-t-1} simplifies further. Let the equilibrium state be described by the PSD $\bmf{f}(x,v,0)=\bmf{f}_{\rm eq}(x,v)=\brvr_{\rm eq}~\gn(v)$, where $\brvr_{\rm eq}$ denotes the homogeneous mass density and $\gn(v)$ represents the velocity distribution such that $\int dv \gn(v)=1$ and $\gn(-v)=\gn(v)$. Using this form of $\bmf{f}(x,v,0)$ in the calculation, as shown in sec.~\eqref{sec:dh_t_a^dr} [see Eqs.~(\ref{mthdr-a-2bis} -\ref{def:Dx})], we get 
\begin{align}
\begin{split}
\msc{C}_{\rm eq}(x_a,u_a,t_a;x_b,u_b,0)= -  \brvr_{\rm eq}\gn(u_b) &\big{[}\delta \mt{g}_{t_a}(x_{\brf(t_a)}(x'_{\bmf{f}(0)}(x)+u_bt_a),u_b) \\
 &~~~~~~- a\brvr_{\rm eq}~\int du~\gn(w)\mt{g}_{n}(x_{\brf(t_a)}(x'_{\bmf{f}(0)}(x)+wt_a),w) \big{]}.
 \end{split}
 \label{ex:corr-ue-t-1bis}
\end{align}
Since in equilibrium the function $ \delta \mt{g}_{t_a}(y,w)$ takes a simpler form $ \delta \mt{g}_{t_a}(y,w)=-\delta(y-x_a)\left[\delta(w-u_a)  - a\brvr_{\rm eq}\gn(u_a) \right]$, on can perform the integral in Eq.~\eqref{ex:corr-ue-t-1bis}. 
The equilibrium space-time correlation is then equal to 
\begin{align}
\begin{split}
\msc{C}_{\rm eq}(x_a,u_a,t_a;&x_b,u_b,0)=\brvr_{\rm eq}(1-a \brvr_{\rm eq}) ~\frac{1}{t_a} \gn \left(\frac{\Delta x}{t_a}\right) \Bigg{\{}\delta\left(u_b-\frac{\Delta x}{t_a}\right) \delta(u_a-u_b)  \cr 
&~~~~~~~~
- a \brvr_{\rm eq} \delta\left(u_b-\frac{\Delta x}{t_a}\right)\gn(u_a) 
- a \brvr_{\rm eq} ~\delta\left(u_a-\frac{\Delta x}{t_a}\right)\gn(u_b) +a^2 \brvr^2_{\rm eq} \gn(u_a) \gn(u_b) \Bigg{\}},
\end{split}
\label{ex:corr-ue-t-2}
\end{align}
where $ \Delta x =  (1-a\brvr_{\rm eq})~(x_a-x_b)$. Integrating the above equation over $u_a$ and $u_b$, we get the equilibrium space-time correlation of the mass density: 
\begin{align}
\msc{C}_{0,0}^{\rm eq}(x_a,t_a;x_b,0)&=\int du_a \int du_b~\msc{C}_{\rm eq}(x_a,u_a,t_a;x_b,u_b,0) \cr
&= \brvr_{\rm eq}(1-a\brvr_{\rm eq})^3 \frac{1}{t_a}\gn \left( \frac{(1-a\brvr_{\rm eq})~(x_a-x_b)}{t_a}\right), \label{eq:S_00(t)-eq}
\end{align}
which again reduces to Eq.~\eqref{fex:s_00(0)} in the limit $t_a \to 0$, assuming $\lim_{t \to 0}(1/t)\gn(x/t)=\delta(x)$. Similarly, we find
\begin{align}
\mc{S}_{m,n}^{\rm eq}(x_a,t_a;x_b,0)=\brvr_{\rm eq}(1-a \brvr_{\rm eq}) ~\frac{1}{t_a} \gn \left(\frac{\Delta x}{t_a}\right) \left[ \left(\frac{\Delta x}{t_a}\right)^m-a\brvr_{\rm eq}\langle u^m\rangle \right]  \left[ \left(\frac{\Delta x}{t_a}\right)^n-a\brvr_{\rm eq}\langle u^n\rangle \right],
\end{align}
for $m,n\ge 1$ where $\langle u^m\rangle = \int du~u^m \gn(u)$.

\subsection{Mass density correlation $\msc{C}_{0,0}(x_a,t_a;x_b,t_b)$:}
\label{sec:corr-mass-den}
The mass density correlation can be obtained by integrating the PSD correlation in Eq.~\eqref{mcS_a_b} over both  the velocities $u_a$ and $u_b$. We have
$\msc{C}_{0,0}(x_a,t_a;x_b,t_b)= -\int du_a \int dv~\delta \mf{f}(x_b,v,t_b) = -\int du_a \delta \varrho(x_b,t_b)$. 
Inserting the form of $\delta \varrho(z,t)$ from Eq.~\eqref{eq:dbrvr(z,t)-fn} and performing some lengthy manipulations, we get  
\begin{align}
\msc{C}_{0,0}&(x_a,t_a;x_b;t_b)= \partial_{x_a} \partial_{x_b} \big{[} (1-a\brvr(x_a,t_a))(1-a\brvr(x_b,t_b)) ~\mc{H}(x_a,t_a;x_b,t_b) \big{]}, \label{eq:S_00(ta,tb)-fn}
\end{align}
where
\begin{align}
\begin{split}
\mc{H}(x_a,t_a;x_b,t_b)
&=\int dy' \int du ~\Theta(x'_a-ut_a-y')\Theta(x'_b-ut_b-y') \brf(y',u,0) \cr 
&
-a\int dy'\int du\int dw ~(1-a\brvr(y,0)) \Theta(x'_a-wt_a-y')\Theta(x'_b-ut_b-y') \cr 
&~\times~~\Big{\{}\frac{2-a\brvr(y,0)}{(1-a\brvr(y,0))^2}\bmf{f}(y,u,0)\bmf{f}(y,w,0)   -\bmf{f}(y,u,0)\brf(x_a',w,t_a)   \cr 
&~~~~~~~~~~~~~
 -\bmf{f}(y,w,0)\brf(x_b',u,t_b) -a \brvr(y,0)\brf(x_a',w,t_a) \brf(x_b',u,t_b) \Big{\}},
 \end{split}
\label{eq:mcA}
\end{align}
where $y=x_{\brf(0)}(y')$, $x'_a=x'_{\bmf{f}(t_a)}(x_a)$,~$x'_b=x'_{\bmf{f}(t_b)}(x_b)$ and $\brf(x'_{\bmf{f}(t)}(x),u,t) = \frac{\bmf{f}(x,u,t)}{1-a\brvr(x,t)}$. It is easy to check that in equilibrium for $t_b=0$, the correlation in the above equation indeed reduces to the equilibrium space-time correlation in Eq.~\eqref{eq:S_00(t)-eq}.

\begin{figure}[t]
\centering
\includegraphics[width=6.5in]{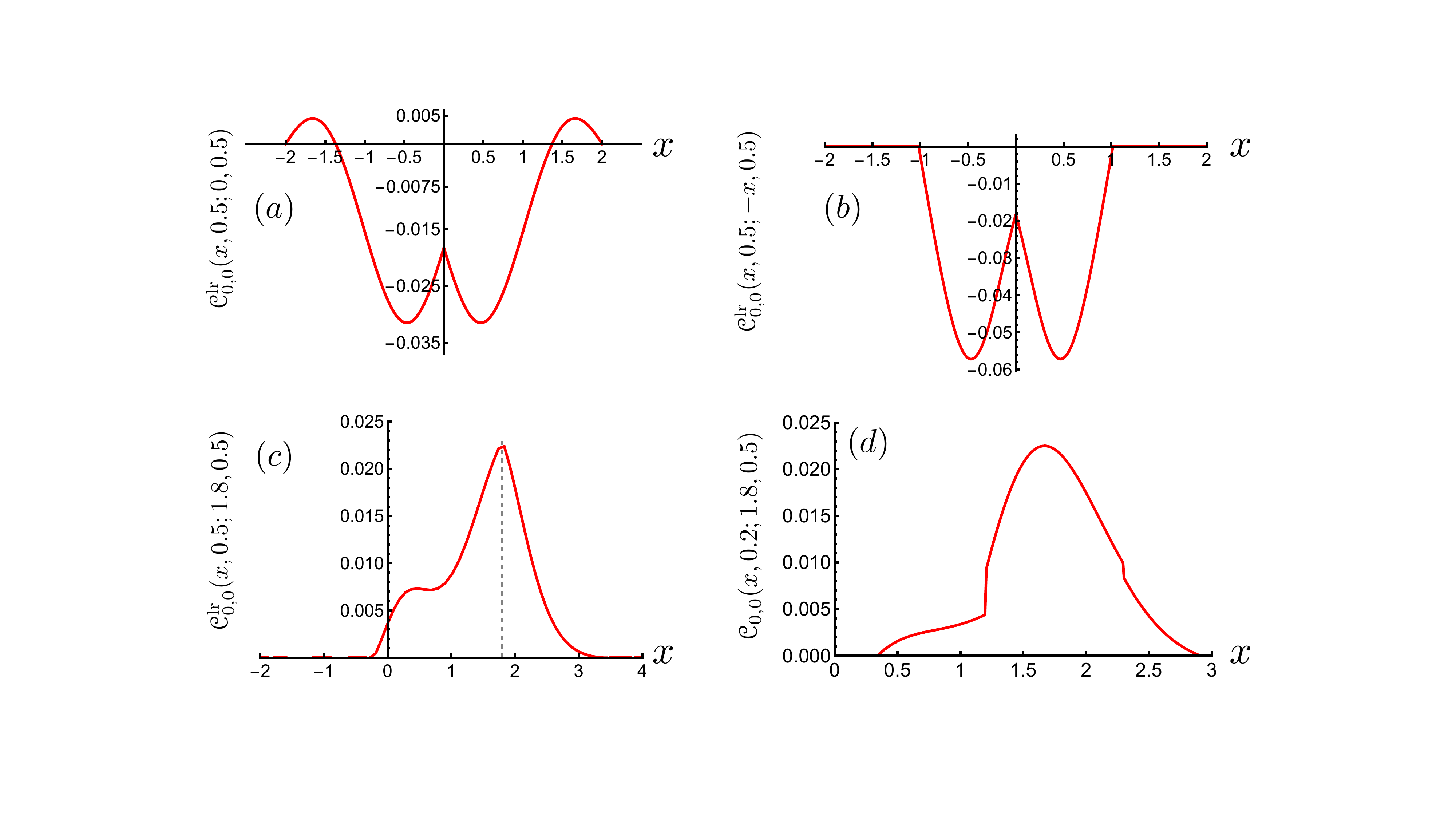} 
\caption{Plot of the correlation $\msc{C}_{0,0}(x_a,t_a;x_b,t_b)$ at different times for the initial condition in Eq.~\eqref{ini-psd}. The plots (a), (b) and (c) correspond to the long-range part of the equal time correlations of mass densities and  the plot in (d) corresponds to unequal space-time correlation {for hard rods of length $a=1$.} Recall, the  densities at two distant locations at a later time $t$ get correlated (long-range correlation) because the fluctuations at these two locations originated from the same initial density fluctuation that coherently got carried to these locations by Euler evolution [see \cite{doyon2023ballistic} for more physical insights]. 
Hence, for the initial condition in Eq.~\eqref{ini-psd}, long-range correlation can get developed only over a finite region over a given duration. The width of the region of course depends on $t_a,~t_b$ and the maximum relative speed of the particles, whereas the location of this region is decided by $x_b$. One finds that the ranges over which the long-range correlations get develop for the four plots are, 
(a) $[-2.002,2.002]$ (b) $[-1.016,1.016]$ (c) $[-0.218,3.359]$  and (d) $[0.342,2.91]$. These ranges can be identified from  the $\Theta$ functions present in the explicit expression in Eqs.~\eqref{eq:S_00(ta,tb)-fn} and \eqref{eq:mcA}. 
We observe  discontinuities in the unequal space time correlation in plot (d) at locations  $x_\pm=x_{\bmf{f}(t_a)}(x'_a(\pm))$ with $x'_a(\pm)=x'_b \pm(t_a-t_b)$ for $x'_b=0.883~ (x_b=1.8),~t_a=0.2$ and $t_b=0.5$. It seems the discontinuities are artifacts of the discrete velocity distribution chosen in Eq.~\eqref{ini-psd}.
}
\label{fig2}
\end{figure}

\paragraph{Equal time mass density correlation:} 
For $t_a=t_b$, the expression of the correlation in Eq.~\eqref{eq:S_00(ta,tb)-fn} is evidently symmetric in $x_a$ and $x_b$ {\it i.e.},
$\msc{C}_{0,0}(x_a,t_a;x_b;t_a)=\msc{C}_{0,0}(x_b,t_a;x_a;t_a)$ and matches exactly with the expression in Eq.~(D.15) of Ref:~\cite{hubner2025diffusive}. 
Taking the derivative explicitly, it is also easy to observe that the equal-time correlation in the above expression has a long-range part $\msc{C}_{0,0}^{\rm lr}$ in addition to a short-range part proportional to $\delta(x_a-x_b)$ :
\begin{align}
\msc{C}_{0,0}(x_a,t_a;x_b;t_a)= (1-a\brvr(x_a,t_a))^2 \brvr(x_a,t_a)\delta(x_a-x_b) + \msc{C}_{0,0}^{\rm lr}(x_a,t_a;x_b;t_a).
\end{align}
As pointed out in Refs.~\cite{doyon2023emergence,doyon2023ballistic}, the long-range part of the correlation survives under three conditions: (i) the initial state must be inhomogeneous, (ii) the system should be interacting, and (iii) there should be at least two hydrodynamic modes with different fluid velocities. It is straightforward to check these conditions from the expression in Eq.~\eqref{eq:S_00(ta,tb)-fn} with $t_a=t_b$. If the initial state is homogeneous in space, then the term inside the curly bracket in Eq.~\eqref{eq:mcA} is zero. Second, if the interaction among the particles is made zero {\it, i.e.}, for $a=0$, only the first term in the expression of $\mc{H}$  in Eq.~\eqref{eq:mcA} survives, again implying no long-range correlation. Checking the third condition requires a little more effort. In this case, we assume $\bmf{f}(y,u,0) = \brvr(y,0)\delta(u-u_0)$. Then $\brf(y',u,0) = \brr(y',0)\delta(u-u_0)$ where $y'=x'_{\bmf{f}(0)}(y)$. Consequently, $\brf(z',u,t_a)=\brf(z'-ut_a,u,0)=\brr(z'-u_0t_a)\delta(u-u_0)$. Using these relations in the expression of $\mc{H}(x_a,t_a;x_b,t_a)$ in Eq.~\eqref{eq:mcA} with $t_a=t_b$ one has 
\begin{align}
\begin{split}
\mc{H}(x_a,t_a;x_b,t_a)
&=(1+a\brf(x'_a-u_0t_a,0))(1+a\brf(x'_b-u_0t_a,0))\cr 
&~~~\times~~\int dy'~\Theta(x'_a-u_0t_a-y')\Theta(x'_b-u_0t_b-y')~\brvr(y,0)(1-a\brvr(y,0)
\end{split}
\label{eq:mcA-1mode}
\end{align}
Inserting this expression of $\mc{H}(x_a,t_a;x_b,t_a)$ in the expression for $\msc{C}_{0,0}(x_a,t_a;x_b;t_a)$ in Eq.~\eqref{eq:S_00(ta,tb)-fn} and using $\brr(x'_a,t_a)=\brr(x'_a-u_0t_a,0)$, we get $\msc{C}_{0,0}(x_a,t_a;x_b;t_a)= (1-a\brvr(x_a,t_a))^2\brvr(x_a,t_a)~\delta(x_a-x_b)$ {\it, that is,} no long-range correlation. 
The above expression clearly shows absence of long-range correlation in case of single hydrodynamic mode and thus verifies the third condition. 

Furthermore, for a symmetric initial condition of the form $\mff(x,v,0)=\varrho(x)\gn(v)$ with $\varrho(-x)=\varrho(x)$ and $\gn(-v)=\gn(v)$, the system is expected to have inversion symmetry even at later times. It is straightforward to see that our expression in Eqs.~\eqref{eq:S_00(ta,tb)-fn} and \eqref{eq:mcA}, indeed possesses the inversion symmetry $\msc{C}_{0,0}(x_a,t_a;x_b;t_a)= \msc{C}_{0,0}(-x_a,t_a;-x_b;t_a)$. These symmetries can easily be demonstrated by numerically evaluating the integrals in Eq.~\eqref{eq:mcA} and plotting the resulting correlation in Eq.~\eqref{eq:S_00(ta,tb)-fn}. For the following initial PSD 
\begin{align}
\mff(y,v,0)=\frac{1+3e^{-y^2}}{3+3e^{-y^2}}~\gn(v),~~\text{with,}~~\gn(v)=\frac{1}{2}\delta(v-1) + \frac{1}{2}\delta(v+1), \label{ini-psd}
\end{align}
we numerically compute the long-range part of the correlation $\msc{C}_{0,0}^{\rm lr}(x_a,t_a;x_b,t_b)$ for different choices of locations and time. We plot these correlations in fig.~\ref{fig2} which demonstrate the symmetry $\msc{C}_{0,0}(x_a,t_a;x_b;t_a)= \msc{C}_{0,0}(x_b,t_a;x_a;t_a)=\msc{C}_{0,0}(-x_a,t_a;-x_b;t_a)$.

\section{Conclusion}
\label{conclusion}
In this paper, we studied the dynamic fluctuations of PSD in a gas of hard rods starting from a slowly varying initial statistical state. To study such fluctuations, recently a general theory called ballistic macroscopic fluctuation theory has been developed for systems supporting ballistic transport. Examples of such systems include certain integrable systems in one dimension, such as a collection of hard rods moving in one dimension in the absence of any external potential for which one can perform explicit calculations. For this system, we review the BMFT by developing the concepts from microscopic description in detail. We found, as in equilibrium systems and in MFT for diffusive systems, that the probability of fluctuations in the PSD on macroscopic space-time scale possesses a large-deviation form with an appropriate large-deviation functional. One important difference between BMFT and MFT is that in the former case one looks at the fluctuations at Euler space-time scaling, whereas in MFT one looks at the diffusive space-time scaling. Hence the source of the noise in the former case is solely from the initial conditions, whereas in the latter it is due to both initial conditions and corse-graining over non-conserved degrees of freedom. 
 
The BMFT can be used to calculate the cumulants and correlations of additive observables through hydrodynamic projections and the principle of {local relaxation} \cite{doyon2023ballistic}. One of the {striking predictions} of this theory is the presence of long-range correlations among conserved densities (and also among local observables) that are established on the Euler scaling limit due to ballistic evolution. In other words, observables in two regions $R$ and $R'$, each of size $\sim \ell$ and separated by distance $\sim \ell$ {get correlated over} time. After time $t \sim \ell$, the correlation between $R$ and $R'$ is also of the order $\ell$ even when there was no correlation initially. We have calculated this correlation between PSDs at two distant locations. We provide an explicit expression of the correlation in an easily computable form and found our result to reproduce the previous computation of the correlation \cite{doyon2023ballistic, hubner2025diffusive}. Such correlations exist when (i) the initial state is inhomogeneous with a length scale of variation $\ell$ (ii) the system is interacting and (iii) the system should have at least two ballistic hydrodynamic modes with different velocities. We explicitly show that our expression indeed becomes local {\it i.e.}, no long-range correlation when any one of the conditions is not satisfied. This long-range correlation is different from what one observes in non-equilibrium stationary states. This correlation is produced by a coherent evolution of initial fluctuation through the Euler hydrodynamic equations. 

Our study can be applied to compute other quantities such as the distribution of net integrated current crossing a location and displacement of a tracer quasiparticle. {In fact a very recent paper has applied this method to compute the full counting statistics in homogeneous state \cite{kethepalli2025ballistic}.} The study of the joint distribution of displacements of two tracers is also an important problem that can be studied using the present approach. The correlation between two tracer rods has been used to develop a fluctuating hydrodynamic theory of hard-rod gas in equilibrium (homogeneous) \cite{ferrari2023macroscopic}. It would be interesting to see the connection between such a fluctuating hydrodynamic equation and BMFT in the context of hard rods and establish a more general fluctuating hydrodynamic theory to study evolution from inhomogeneous initial states. Due to the general structure of BMFT, it would be interesting to apply this theory to other integrable systems such as the Toda system for which GHD has been established \cite{doyon2019generalized, spohn2021hydrodynamic}. Finally, it would be interesting to explore the possibility of extending the BMFT theory for hard rods in the presence of slowly varying external potentials.

\section{Acknowledgement}
The author thanks Subhro Bhattacharjee, Sthitadhi Roy, and Abhishek Dhar for insightful discussions. I would also like to acknowledge the financial support from DST, Government of India Grant under Project No. CRG/2021/002455 and the MATRICS grant MTR/2021/000350 from the ANRF, DST, Government of India, and also the support from the DAE, Government of India, under Project No. RTI4001.

\appendix
\addtocontents{toc}{\fixappendix}
\section{Entropy of hard rods inside a box}
\label{ent-hr}
As shown in the schematic diagram in Fig.~\ref{schema}, the entire system is divided into blocks of size $\Delta \mtx \sim \ell$, labeled by the index $\mt{Y}_k$ with $k=...-1,0,1,...$. We focus on the $k^{\rm th}$ block that has $N_k$ number of particles having velocities in the range $-\infty < v<\infty$. The phase space corresponding to this block is $(\mtx,v) \in [0,\ell]\times [-\infty,\infty]$. We further divide this phase space into small ranges $\Delta v$ in the velocity direction and distribute the $N_k$ number of particles into these velocity cells such that $\sum_vN_{v|k}=N_k$. To compute the Boltzmann entropy corresponding to the macrostate specified by $\hat{M}=\{N_{v|k}\}$, one needs to compute the phase space volume $\big{|}\Gamma_{\hat{M}}\big{|}$ associated with this macrostate $\hat{M}$, which is formally given by ({see} \cite{chakraborti2022entropy,pandey2023boltzmann} for the computation of the Boltzmann entropy of classical and quantum ideal gases)
\begin{align}
\big{|}\Gamma_{\hat{M}}\big{|} = \frac{N_k!}{\prod_v N_{v|k}!} \prod_{v} (\Delta v)^{N_{v|k}} ~\big{|}\Gamma_{N_k}^{\Delta \mtx}\big{|}, \label{Gamma_M-1}
\end{align}
where 
\begin{align}
|\Gamma_{n}^\ell \big{|} &= \int_{(n-1/2)a}^{\ell -a/2}d\mtx_n...\int_{3a/2}^{\mtx_3-a}d\mtx_2 \int_{a/2}^{\mtx_2-a}d\mtx_1~~1,\cr
&=\int_{0}^{\ell -na}d\mt{z}_n...\int_{0}^{\mt{z}_3}d\mt{z}_2 \int_{0}^{\mt{z}_2}d\mt{z}_1~1,~~\left[ \text{using}~\mt{z}_j=\mt{x}_j-(j-1/2)a\right], \cr
&=\frac{1}{n!} \prod_{j=1}^n~\int_0^{\ell -na}d\mt{z}_j = \frac{(\ell-na)^n}{n!}. \label{Gamma_n}
\end{align}
Inserting this expression in Eq.~\eqref{Gamma_M-1}, we get 
\begin{align}
\big{|}\Gamma_{\hat{M}}\big{|} &=  N_k! \prod_v \frac{(\Delta v)^{N_{v|k}}}{N_{v|k}!}~\frac{(\Delta \mtx-N_ka)^{N_k}}{N_k!} 
= \prod_v \frac{\left\{\Delta v (\Delta \mtx - aN_k)\right\}^{N_{v|k}}}{N_{v|k}!}.
\end{align}
The entropy is then given by $S_k[\{N_{v|k}\}] = \ln \big{|}\Gamma_{\hat{M}}\big{|}$ (assuming the Boltzmann constant $K_B=1$). Approximating the factorials using the Stirling approximation for large $N_{v|k}$, one finds
\begin{align}
\begin{split}
S_k[\{N_{v|k}\}] &\approx - \sum_v ~N_{v|k}\ln \left( \frac{N_{v|k}}{\Delta v(\ell-a N_k)}\right)\cr
&=-N_k \sum_v \Delta v~\mff_{v|k} \ln \left( \frac{\mff_{v|k}N_k }{\ell-a N_k}\right),~~\text{with}~N_{v|k}=\Delta v~\mff_{v|k}N_k, \cr
&=-\ell \sum_v \Delta v~\mff_{v|k} \varrho_k \ln \left( \frac{\mff_{v|k}\varrho_k }{1-a \varrho_k}\right),
\end{split}
\label{derv:S_k-a}
\end{align}
where we have used $N_k=\varrho_k \ell$.

\section{Derivation of the saddle-point equations Eqs.~(\ref{eq:sad-1}) - (\ref{eq:sad-4})}
\label{derv:sad-eq}
Here we derive the saddle-point equation for a general observable $\mc{A}=\int_0^{\tilde{T}}dt \int dx \int dv{\msc{A}(}x,v,t)\mff(x,v,t) $. For that, we consider the action
\begin{align}
\mc{S}^{\mc{A}}_{\lambda}[\fn, \hn]=  \msc{S}_\mu[\fn,\hn]+ \lambda \int_0^{\tilde{T}}dt\int dx'\int dv~{\msc{A}(x}'+a\Fn(x',t),v,t)\fn(x',v,t).
\label{def:mcS-p-a}
\end{align}
Note, for the action $\mc{S}^{\mt{f}_{\rm r}}_{\lambda}$ in Eq.~\eqref{def:mcS-p} one needs to choose 
\begin{align}
{\msc{A}(x},v,t)=\delta(x-x_a)\delta(v-u_a)\delta(t-t_a). \label{app:A-for-C}
\end{align} 
We consider the variation
$\delta\mc{S}^{\mc{A}}_\lambda=\mc{S}^{\mc{A}}_\lambda[\fn_\lambda^{\rm sd}+\delta \fn, \hn_\lambda^{\rm sd}+\delta \hn, \mu_{\rm sd}+\delta \mu] - \mc{S}^{\mc{A}}_\lambda[\fn_\lambda^{\rm sd}, \hn_\lambda^{\rm sd}, \mu_{\rm sd}]$. Equating this variation to zero, one gets the saddle-point equations. It is straightforward to show that 
\begin{align}
\delta\msc{S}^{\mc{A}}_\lambda[\fn_\lambda^{\rm sd},&\delta \fn]=\int d\gamma' {\delta\fn(0)\left[\mathcal{R}_{\fn_\lambda^{\rm sd}(0)}[\psi(\gamma')]+\ln \fn_\lambda^{\rm sd}(0) + \mu \right] }+\delta \mu \left(\ell \int d\gamma' \fn_\lambda^{\rm sd}(0)-N\right), \cr
&+\lambda \int_0^Tdt \int d\gamma' ~\mathcal{R}_{\fn_\lambda^{\rm sd}(t)}[{\msc{A}(x}_{\fn_\lambda^{\rm sd}(t)}(x'),v,t)]\delta \fn(t) + \int_0^Tdt \int d\gamma'~\delta \hn(t)\left[ \partial_t \fn_\lambda^{\rm sd}+v\partial_{x'}\fn_\lambda^{\rm sd}\right] , \cr 
&+\int d\gamma'~{\left[ \hn_\lambda(\gamma',T) \delta \fn(T) - \hn_\lambda(\gamma',0)\delta \fn(0)\right]} ,-\int_0^T\int d\gamma'~\delta \fn(t)\left[ \partial_t \hn_\lambda^{\rm sd}+v\partial_{x'}\hn_\lambda^{\rm sd}\right],  
\end{align}
where $\gamma'=(x',v)$, $d\gamma'=dx' dv$ and $\mathcal{R}_{\mff(t)}[H(x,v)]$ is defined in Eq.~\eqref{R[H]}. Although computing the variations for other terms is straightforward, we present only the details for the term
\begin{align}
\mathcal{T}^{\mc{A}}[\fn(t)]=\lambda \int_0^T\int dx' \int dv \fn(x',v,t)~{\msc{A}(x}'+a\Fn(x',t),v,t).
\end{align}
The variation of $\mc{T}^A$ is 
\begin{align}
\delta \mathcal{T}^{\mc{A}}[\fn_\lambda^{\rm sd} , \delta \fn] &= \lambda \int_0^Tdt\int dx' \int dv~ \delta \fn(x',v,t)~{\msc{A}(x}_{\fn_\lambda^{\rm sd}(t)}(x'),v,t)\cr
& ~~~~~~~~~~+ \lambda \int_0^Tdt \int dx'\int dv~ \delta \fn(x',v,t)  \cr 
&~~~~~~~~~~~~~~~~~\times~\int dy' \Theta(y'-x')\int du ~\fn_\lambda^{\rm sd}(y',u,t) \partial_{x_{\fn_\lambda^{\rm sd}(t)}(y')}{\msc{A}(x}_{\fn_\lambda^{\rm sd}(t)}(y'),u,t) \cr 
&=\lambda \int_0^Tdt \int dx' \int dv~ \delta \fn(x',v,t)~\Big{[}{\msc{A}(}x_{\fn_\lambda^{\rm sd}(t)}(x'),v,t) \cr 
&~~~~~~~~~~~~~~~~~
+ \int dy\int du~\Theta(y-x_{\fn_\lambda^{\rm sd}(t)}(x')) \mf\fn_\lambda^{\rm sd}(y,u,t) \partial_{y}A(y,u,t) \Big{]} \cr
&=\lambda \int_0^Tdt\int dx' \int dv~ \delta \fn(x',v,t)~\mathcal{R}_{\fn_\lambda^{\rm sd}(t)}[{\msc{A}(x}_{\fn_\lambda^{\rm sd}(t)}(x'),v,t)].
\end{align}
Equating $\left(\frac{\delta \mc{S}^{\mc{A}}_\lambda}{\delta \fn}\right)_{\fn=\fn_\lambda^{\rm sd}}=0$,  $\left(\frac{\delta \mc{S}^{\mc{A}}_\lambda}{\delta \hn}\right)_{\hn=\hn_\lambda^{\rm sd}}=0$, and $\left( \frac{\partial \mc{S}^{\mc{A}}_\lambda}{\partial \mu}\right)_{\mu = \mu_{\rm sd}}=0$ with $A$ given in Eq.~\eqref{app:A-for-C}, one gets the saddle-point equations in Eq.~\eqref{eq:saddle}.

\subsection{Proof of Eq.~\eqref{eq-to-find-mfrkf(0)}}
\label{derv:f(x,v,0)}
For convenience, we start by rewriting Eq.~\eqref{eq:sad-4-dc-1} 
\begin{align}
\mathcal{R}_{\mff_\lambda(0)}[\psi(x_{\fn_\lambda(0)}(x'),v)]+\ln \fn_\lambda(x',v,0)+\mu =\lambda~\delta \mt{g}_0(x',v),\label{eq:sad-4-dc-1-a}
\end{align}
where, $x'=x'_{\mff_\lambda(0)}(x)$ and recall from Eq.~\eqref{h(0)} that $\delta \mt{g}_0(x',v)=\delta \mt{g}_{t_a}(x_{\fn_\lambda(t_a)}(x'+vt_a),v)$.
{We first note that}
\begin{align}
\partial_{x'}\mathcal{R}_{\mff_\lambda(0)}[\psi(x_{\fn_\lambda(0)}(x'),v)]&=(1+a \rho(x',0))~\partial_{x_{\fn_\lambda(0)}(x')}\psi(x_{\fn_\lambda(0)}(x'),v) \cr
&~~~~~~~~~-a\int du \left\{ \partial_{x_{\fn_\lambda(0)}(x')} \psi(x_{\fn_\lambda(0)}(x'),u)\right\}\fn_\lambda(x',u,0),
\end{align}
{which implies}
\begin{align}
\int dv  \big{\{} \partial_{x_{\fn_\lambda(0)}(x')} \psi(x_{\fn_\lambda(0)}(x'),&v)\big{\}} \fn_\lambda(x',v,0) = \int dv~\fn_\lambda(x',v,0)~\partial_{x'}\mathcal{R}_{\mff_\lambda(0)}[\psi(x_{\fn_\lambda(0)}(x'),v)].
\end{align}
Now using Eq.~\eqref{eq:sad-4-dc-1-a} on the right-hand side of the above equation we get 
\begin{align}
\begin{split}
\int dv  \big{\{} \partial_{x_{\fn_\lambda(0)}(x')} \psi(x_{\fn_\lambda(0)}(x'),&v)\big{\}} \fn_\lambda(x',v,0) \cr
&=- \int dv ~\partial_{x'}\fn_\lambda(x',v,0)+\lambda \int dv~\fn_\lambda(x',v,0)\partial_{x'} \delta\mt{g}_0(x',v).
\end{split}
\label{identity-1}
\end{align}
To proceed, we first convert $\partial_{x'} ( \bullet)= (1+a \rho(x',0))\partial_{x_{\fn_\lambda(0)}(x')} ( \bullet)$. For example, we write 
\begin{align}
\begin{split}
\int dv~ \partial_{x'}\fn_\lambda(x',v,0) &= \int dv~ (1+a\rho(x',0))~\partial_{x_{\fn_\lambda(0)}(x')} \fn_\lambda(x',v,0)\cr 
&=\int dv~\fn_\lambda(x',v,0) ~\partial_ {x_{\fn_\lambda(0)}(x')}[\ln \fn_\lambda(x',v,0) + a \rho(x',0)].
 \end{split}
\label{identity-2}
\end{align}
Similarly, we have 
\begin{align}
\int dv~\fn_\lambda(x',v,0)\partial_{x'} \delta \mt{g}_0(x',v) &= \int dv~\fn_\lambda(x',v,0)(1+a\rho(x',0))~\partial_{x_{\fn_\lambda(0)}(x')}\{\delta \mt{g}_0(x',v)\} \cr
\begin{split}
&=\int dv~\fn_\lambda(x',v,0) \partial_{x_{\fn_\lambda(0)}(x')}~[\delta \mt{g}_0(x',v)] \cr 
&~~~~+ {a}\int dv~\fn_\lambda(x',v,0) \left \{ \int du~\fn_\lambda(x',u,0) \partial_{x_{\fn_\lambda(0)}(x')}[\delta \mt{g}_0(x',u) ] \right \} 
 \end{split}
 \label{identity-3}
\end{align}
Inserting the expressions from Eqs.~\eqref{identity-2} and \eqref{identity-3}, on the right-hand side of Eq.~\eqref{identity-1}, we get 
\begin{align}
\int dv ~\fn_\lambda(x',v,0)&~{\partial_{x_{\fn_\lambda(0)}(x')} \Big{\{}  \psi(x_{\fn_\lambda(0)}(x'),v)+\ln \fn_\lambda(x',v,0) + a \rho(x',0) \big{\}} }\cr 
&- \lambda~\partial_{x_{\fn_\lambda(0)}(x')}~[\delta \mt{g}_0(x',v)]-{a}\lambda \int du~\fn_\lambda(x',u,0) \partial_{x_{\fn_\lambda(0)}(x')} [\delta \mt{g}_0(x',u)] \Big{\}}=0
\label{identity-4}
\end{align}
Since $\fn_\lambda(x',v,0)$ is an arbitrary positive density function, we must equate the term inside $\{..\}$ in the above equation to zero {\it i.e.}, we have 
\begin{align}
\partial_{x} \Big\{ \psi(x_{\fn_\lambda(0)}(x'_{\mff(0)}(x),v)\big{\}} &+\ln \fn_\lambda(x'_{\mff(0)}(x),v,0) + a \rho(x'_{\mff(0)}(x),0) 
-\lambda \delta  \mt{g}_0(x'_{\mff(0)}(x),v)\Big{\}} \cr 
&={a}\lambda \int du~\fn_\lambda(x'_{\mff(0)}(x),u,0) \partial_{x} [\delta \mt{g}_0(x'_{\mff(0)}(x),u) ], \
\end{align}
where we have used the transformation $x'=x'_{\mff(0)}(x)$.
Integrating over $x$ and reintroducing the constant $\mu$ as the integration constant, we get 
\begin{align}
\begin{split}
\psi(x,v) + \ln \fn_\lambda(x'_{\mff(0)}(&x),v,0) +a \rho(x'_{\mff(0)}(x),0) +\mu= \cr
&\lambda \Big{[} \delta \mt{g}_0(x'_{\mff(0)}(x),v) -a\int dy~\Theta(y-x)~ \int du~\fn_\lambda(x'_{\mff(0)}(y),u,0)\partial_{y}\delta \mt{g}_0(x'_{\mff(0)}(y),u) \Big{]}.
\end{split}
\label{identity-5}
\end{align}
This equation can conveniently be written in terms of hard-rod PSD $\mff(x,v,0)$ as in Eq.~\eqref{eq-to-find-mfrkf(0)}. This relation can also be proved using the fact that the operator $\msc{R}_{\mff(t)}$ is the inverse of the
operator $\mc{R}_{\mff(t)}$ {\it i.e.}, $\msc{R}_{\mff(t)}[\mc{R}_{\mff(t)}[H(x,v)]] =H(x,v)$. This fact is proved in~\ref{sec:mscRmcR=I}. Recalling the definition of $\msc{R}_{\mff(t)}[H(x,v)]$ in Eq.~\eqref{Rsc[H]}, it is easy to show that $\msc{R}_{\mff_\lambda(t)}[\mu]=\mu$ and 
\begin{align}
\msc{R}_{\mff_\lambda(t)}[\ln \fn_\lambda(x',v,0)] = \ln \fn_\lambda(x',v,0) + a\rho(x',0),
\end{align}
where $x'=x'_{\mff_\lambda(0)}(x)$. Hence, applying $\msc{R}_{\mff_\lambda(t)}$ on both sides of Eq.~\eqref{eq:sad-4-dc-1-a} and using the above relations, we immediately get 
\begin{align}
\psi(x,v) + \ln \fn_\lambda(x'_{\mff(0)}(&x),v,0) +a \rho(x'_{\mff(0)}(x),0) +\mu=\lambda \msc{R}_{\mff_\lambda(0)}[\mt{g}_{0}(x'_{\mff(0)}(x),v)].
\end{align}

\subsection{Proof of $\msc{R}_{\mff(t)}[\mc{R}_{\mff(t)}[H(x,v)]]=H(x,v)$}
\label{sec:mscRmcR=I}
We start by rewriting the definitions of the operators $\mc{R}_{\mff(t)}$ and $\msc{R}_{\mff(t)}$ from Eq.~\eqref{R[H]} and Eq.~\eqref{Rsc[H]}, respectively, 
\begin{align}
\mathcal{R}_{\mff(t)}[H(x,v)]
&=H(x,v) +a \int dy\int du\left\{ \partial_{y} H(y,u)\right\}\mathfrak{f}(y,u,t)\Theta(y-x). 
\label{R[H]-a}
\end{align}
and 
 \begin{align}
\mathscr{R}_{\mff(t)}[H(x,v)]
&=H(x,v) -a \int dy~\Theta(y-x)~\left[ \int du\left\{ \partial_{y} H(y,u)\right\}\frac{\mff(y,u,t)}{1-a\varrho(y,t)}\right].
\label{Rsc[H]-a}
\end{align}
We evaluate 
 \begin{align}
\msc{R}_{\mff(t)}[\mc{R}_{\mff(t)}[H(x,v)]]
&=\mc{R}_{\mff(t)}[H(x,v)] -a \int dy~\Theta(y-x)~\left[ \int du\left\{ \partial_{y} \mc{R}_{\mff(t)}[H(y,u)]\right\}\frac{\mff(y,u,t)}{1-a\varrho(y,t)}\right],  \displaybreak[3]\notag \\
&= \mc{R}_{\mff(t)}[H(x,v)] -a \int dy~\Theta(y-x)~ \int du~\frac{\mff(y,u,t)}{1-a\varrho(y,t)}~\big{[} \partial_yH(y,u)  \displaybreak[3]\notag\\
&~~~~~~~~~~~~~~
-a\int dz \int du'~\delta (z-y)\left\{ \partial_{z} H(z,u')\right\}\mathfrak{f}(z,u',t) \big{]}  \displaybreak[3]\notag \\
&= H(x,v)+a \int dy~\Theta(y-x) \int du~\left\{ \partial_{y} H(y,u)\right\}\mathfrak{f}(y,u,t),  \displaybreak[3] \notag \\
&-a \int dy~\Theta(y-x)~ \int du~\frac{\mff(y,u,t)}{1-a\varrho(y,t)}~\big{\{} \partial_yH(y,u)
-a\int du'~\left\{ \partial_{y} H(y,u')\right\}\mathfrak{f}(y,u',t) \big{\}},  \displaybreak[3] \notag  \\
&= H(x,v)+a \int dy~\Theta(y-x) \int du~\left\{ \partial_{y} H(y,u)\right\}\mathfrak{f}(y,u,t)  \displaybreak[3] \notag \\
&-a \int dy~\Theta(y-x)~(1+a\rho(x'_{\mff(t)}(y),t)) \int du~\mff(y,u,t)~ \partial_yH(y,u)  \displaybreak[3] \notag \\
&+a^2 \int dy~\Theta(y-x)~\frac{\varrho(y,t)}{1-a\varrho(y,t)}\int du'~\left\{ \partial_{y} H(y,u')\right\}\mathfrak{f}(y,u',t), \displaybreak[3] \notag \\
&= H(x,v)+a \int dy~\Theta(y-x) \int du~\left\{ \partial_{y} H(y,u)\right\}\mathfrak{f}(y,u,t)   \displaybreak[3] \notag\\
&-a \int dy~\Theta(y-x)~(1+a\rho(x'_{\mff(t)}(y),t)) \int du~\mff(y,u,t)~ \partial_yH(y,u)  \displaybreak[3] \notag \\
&+a^2 \int dy~\Theta(y-x)~\rho(x'_{\mff(t)}(y),t)\int du'~\left\{ \partial_{y} H(y,u')\right\}\mathfrak{f}(y,u',t),  \displaybreak[3] \notag \\
&= H(x,v).
\label{Rsc[H]-a-prf}
\end{align}

\section{Solution of the saddle-point equations in hard-rod picture}
\label{derv:sad-eq-hr}
Here we solve the saddle-point equations \eqref{eq:saddle-a} without going to the hard point picture.
Once again for notational simplicity, we, from now on omit the superscript `sd' from the saddle-point functions. We also omit the subscript $`\lambda'$ for the same reason in this section. 
In order to solve Eqs.~\eqref{eq:saddle-a}, it seems convenient to imagine an infinite-dimensional vector space in $v$ at each point $(x,t)$ 
and introduce the bra-ket notation to define the following functions
\begin{align}
\begin{split}
\mathfrak{f}(x,v,t) &\equiv \braket{v}{\mathfrak{f}(x,t)}, \\
\mf{h}(x,v,t) &\equiv \braket{v}{\mf{h}(x,t)}.
\end{split}
\end{align}
The inner product in this space is defined as 
\begin{align}
\braket{\mf{g}_1(x,t)}{\mf{g}_2(x,t)}=\int dv~\mf{g}_1(x,v,t)\mf{g}_2(x,v,t).
\end{align}
Linear operators are defined as (infinite-dimensional) matrices. We define the following particular operator
\begin{align}
\bra{v}\mathbb{A}[\mf{f}](x,t)\ket{u} = \mathbb{A}_v^{u}[\mf{f}] &= \delta(v-u)v_{\rm eff}(x,v,t) + a\left[ v_{\rm eff}(x,v,t)-u\right] \frac{\mathfrak{f}(x,v,t)}{1-a\varrho(x,t)}, \cr
&=\delta(v-u)v_{\rm eff}(x,v,t) + a\left[ v_{\rm eff}(x,v,t)-u\right] ~\mf{f}_n(x,v,t), \label{eq:mbA-1-a}
\end{align}
where we define the normal PSD 
\begin{align}
\mf{f}_n(x,v,t)= \frac{\mathfrak{f}(x,v,t)}{1-a\varrho(x,t)}. \label{eq:f_n-f-a}
\end{align}
Using the vector notation, we write the equations (\ref{eq:sad-2-a}) and (\ref{eq:sad-1-a}) as
\begin{align}
&\partial_t \ket{\mathfrak{f}(x,t)} + \mathbb{A}(x,t)\partial_{x} \ket{\mathfrak{f}(x,t)} =0, \label{eq:vec-sad-f-a}\\
&\partial_t \ket{\mf{h}(x,t)} + \mathbb{A}^T(x,t)\partial_{x} \ket{\mf{h}(x,t)} =\lambda \ket{{\msc{A}(x},t)},
\label{eq:vec-sad-h-a}
\end{align}
where $\mathbb{A}^T$ represents the transpose of the matrix $\mathbb{A}$ and $\braket{v}{{\msc{A}(x},t)} ={\msc{A}(x},v,t) = \mf{b}(x,v)\delta(t-T)$.
One needs to solve these equations with boundary conditions
\begin{align}
\begin{split}
\ket{\mathfrak{f}(x,t)}_{x \to \pm \infty} &=\ket{\mathfrak{f}(x,0)}_{x \to \pm \infty} ,\\
\ket{\mf{h}(x,t)}_{x \to \pm \infty} &=0,
\end{split}
\label{bc-vec-a}
\end{align}

\subsection{Solving the Eqs.~ \eqref{eq:vec-sad-f-a} and \eqref{eq:vec-sad-h-a} }
To solve the Eq.~\eqref{eq:vec-sad-f-a}, one needs to diagonalise the matrix $\mathbb{A}$ given in Eq.~\eqref{eq:mbA-1-a}.  For this we define the (infinite dimensional) matrix operator 
\begin{align}
\mb{T}=-a\ket{\varsigma_0}\bra{\varsigma_0},
\end{align}
where $\braket{v}{\varsigma_0}=1,~~\forall v$. It is easy to see that the mass density  can be written as $\varrho(x,t)=\braket{\varsigma_0}{\mathfrak{f}(x,t)}$. Consequently the normal PSD defined in Eq.~\eqref{eq:f_n-f-a} can be written in the following vector notation
\begin{align}
\ket{\mf{f}_n(x,t)}=\frac{\ket{\mathfrak{f}(x,t)}}{1-a\braket{\varsigma_0}{\mathfrak{f}(x,t)}}. \label{def:vec-f_n-a}
\end{align}
We are now in a position to diagonalise the matrix $\mb{A}$. By explicit calculation, as will be shown below, the matrix $\mb{A}$ can be diagonalised as 
\begin{align}
\mb{R}\mb{A}\mb{R}^{-1} = \mb{V}_{\rm eff}, \label{eq:mbA-diag}
\end{align}
where
\begin{align}
\mb{R}(x,t)&=1+a\ket{\mf{f}_n(x,t)}\bra{\varsigma_0}, \label{def:mbR-a}
\end{align}
and
\begin{align}
\bra{v}\mb{V}_{\rm eff}(x,t)\ket{u}&=v_{\rm eff}(x,v,t)\delta(v-u). \label{def:mbV-a}
\end{align}
It is easy to check that $\mb{R}^{-1}=1-a\ket{\mathfrak{f}(x,t)}\bra{\varsigma_0}$ as follows:
\begin{align}
\mb{R}^{-1}&=(1+a\ket{\mf{f}_n}\bra{\varsigma_0})^{-1}=1-a\ket{\mf{f}_n}\left( 1-a\bra{\varsigma_0}\ket{\mf{f}_n} +a^2 \bra{\varsigma_0}\ket{\mf{f}_n}^2 +...\right) \bra{\varsigma_0} \cr 
&=1-\frac{a\ket{\mf{f}_n}\bra{\varsigma_0}}{1+a\braket{\varsigma_0}{\mf{f}_n}} = 1-a\ket{\mff}\bra{\varsigma_0},
\end{align}
where we have used Eq.~\eqref{def:vec-f_n-a} and 
\begin{align}
1-a \bra{\varsigma_0}\ket{\mathfrak{f}(x,t)} = \frac{1}{1+a\bra{\varsigma_0}\ket{\mf{f}_n(x,t)}}. \label{rel:rho-rho_n-a}
\end{align}
Also note that using the above equation and the definition of $\mb{R}$ in Eq.~\eqref{def:mbR-a}, one can rewrite Eq.~\eqref{def:vec-f_n-a} as 
\begin{align}
\ket{\mf{f}_n(x,t)}=\mb{R}\ket{\mathfrak{f}(x,t)}. \label{def:vec-f_n-2-a}
\end{align}
Now we check
\begin{align}
\bra{v}\mb{R}\mb{A}\mb{R}^{-1} \ket{u} &= \bra{v}\mb{A}\ket{u}+a\mf{f}_n(v)\int dw~ \bra{w}\mb{A}\ket{u} -a \bra{v}\mb{A}\ket{f} -a^2\mf{f}_n(v) \bra{\varsigma_0}\mb{A}\ket{f}, \label{mbA-diag-1-a}
\end{align}
where we have used $\braket{v}{\varsigma_0}=1,~~\forall v$ and completeness relation $\int dw~\ket{w}\bra{w}=\mb{I}$.
It is easy to show that
\begin{align}
\int dv~v_{\rm eff}(v)\mathfrak{f}(v)=\varrho(x,t)\vartheta(x,t) =\mc{J}(x,t). \label{rel-0-a}
\end{align}
where we have used the expression of $v_{\rm eff}$ given in Eq.~\eqref{def:v_eff} along with the definitions of the mass density $\varrho(x,t)$ and flow velocity $\vartheta(x,t)$  given in Eq.~\eqref{def:rho-F} and \eqref{eq:flow-vel-a} respectively. Here $\mc{J}(x,t)=\varrho(x,t)\vartheta(x,t)$ denotes the current of the mass flow.
Using the integral in Eq.~\eqref{rel-0-a} one can show that
\begin{align}
\bra{\varsigma_0}\mb{A}\ket{f}&=\int dv' \int du'~\left[\delta(v'-u')v_{\rm eff}(v')\mathfrak{f}(u')+a\mf{f}_n(v')\left(v_{\rm eff}(v')-u'\right)\mathfrak{f}(u') \right] = \varrho(x,t)\vartheta(x,t),
\label{rel-1-a}
\end{align}
\begin{align}
\int dv' \bra{v'}\mb{A}\ket{u} &= \int dv'~\left[ v_{\rm eff}(v')\delta(v'-u) + a \mf{f}_n(v')\left( v_{\rm eff}(v')-u\right)\right] \cr
&=v_{\rm eff}(u) + a\frac{[\vartheta(x,t) \varrho(x,t) -u\varrho(x,t)]}{1-a\varrho(x,t)}, \label{rel-2-a}
\end{align}
and
\begin{align}
\bra{v}\mb{A}\ket{\mf{f}}=\int dv'  \bra{v}\mb{A}\ket{v'}\mathfrak{f}(v') = v_{\rm eff}(v)\mathfrak{f}(v) +a\mf{f}_n(v)[v_{\rm eff}(v)\varrho(x,t) - \vartheta(x,t) \varrho(x,t)]. \label{rel-3-a}
\end{align}
Using the relations in Eqs.~\eqref{rel-1-a} - \eqref{rel-3-a} in Eq.~\eqref{mbA-diag-1-a}, one can straightforwardly 
show the the matrix $\mb{R}$ can indeed diagonalise the matrix $\mb{A}$ {\it i.e.}, we have 
\begin{align}
\bra{v}\mb{R}\mb{A}\mb{R}^{-1} \ket{u} &=\delta(v-u) v_{\rm eff}(x,v,t). \label{mbA-diagonalised-a}
\end{align}
as announced in Eq.~\eqref{eq:mbA-diag} in matrix notation. Taking transpose of this equation we also have 
\begin{align}
\mb{R}^{-T}\mb{A}^T\mb{R}^{T}=\mb{V}_{\rm eff}, \label{eq:mbA-diag-tran}
\end{align}
where $\mb{R}^{-T}$ is shorthand notation for $\left(\mb{R}^{-1}\right)^T$. One can easily show that
\begin{align}
\mb{R}^T=1+a\ket{\varsigma_0}\bra{\mf{f}_n},~~\text{and}~~\mb{R}^{-T}=1-a\ket{\varsigma_0}\bra{\mf{f}}. \label{eq:R^T-Rinv^T}
\end{align}
Using the matrix operators $\mb{R}$ and $\mb{R}^T$ we diagonalise the equations \eqref{eq:vec-sad-f-a} and \eqref{eq:vec-sad-h-a}. We first apply $\mb{R}$ on both sides of Eq.~\eqref{eq:vec-sad-f-a}. From the transformation in Eq.~\eqref{def:vec-f_n-2-a}, we write 
\begin{align}
\begin{split}
\mb{R}\partial_t \ket{\mathfrak{f}(x,t)}&=\partial_t\ket{\mf{f}_n(x,t)} - \left( \partial_t \mb{R}\right) \ket{\mathfrak{f}(x,t)},\\
\mb{R}\partial_{x} \ket{\mathfrak{f}(x,t)}&=\partial_{x}\ket{\mf{f}_n(x,t)} - \left( \partial_{x} \mb{R}\right) \ket{\mathfrak{f}(x,t)},
\end{split}
\label{rel:partial-f-f-n}
\end{align}
using which,  we see Eq.~\eqref{eq:vec-sad-f-a} gets transformed to a normal mode equation
\begin{align}
\partial_t \ket{\mf{f}_n(x,t)} + \mb{V}_{\rm eff}(x,t)\partial_{x} \ket{\mf{f}_n(x,t)} =0, \label{eq:vec-sad-f_n-a}
\end{align}
where $\ket{\mf{f}_n(x,t)}$ is the normal mode vector [as was mentioned previously in Eq.~\eqref{eq:f_n-f-a}] and  the diagonal matrix $\mb{V}_{\rm eff}$ is defined in Eq.~\eqref{def:mbV-a}. One has to solve this equation with boundary condition $\ket{\mf{f}_n(x,t)}_{x \to \pm \infty}=0$. Note Eq.~\eqref{eq:vec-sad-f_n-a} straightforwardly implies
\begin{align}
\partial_t \mb{R}(x,t)+ \mb{V}_{\rm eff}\partial_{x}\mb{R}(x,t) =0. \label{eq:sad-mbR}
\end{align}
We now turn our attention to Eq.~\eqref{eq:vec-sad-h-a}. Applying $\mb{R}^{-T}$ on both sides of this equation we get
\begin{align}
\mb{R}^{-T}\left(\partial_t \ket{\mf{h}(x,t)}\right) + \mb{V}_{\rm eff} \mb{R}^{-T}\left(\partial_{x} \ket{\mf{h}(x,t)}\right) = \lambda \mb{R}^{-T}\ket{{\msc{A}(x},t)}.
 \label{eq:h-dig-1-a}
\end{align}
It is easy to show that the following vectors 
\begin{align}
\ket{\mf{g}^t_n(x,t)}&= \mb{R}^{-T}\left(\partial_t \ket{\mf{h}(x,t)}\right) \\
\ket{\mf{g}^x_n(x,t)}&= \mb{R}^{-T}\left(\partial_{x} \ket{\mf{h}(x,t)}\right). 
\end{align}
satisfy 
\begin{align}
\partial_{x}\ket{\mf{g}^t_n(x,t)} = \partial_t\ket{\mf{g}^x_n(x,t)}. \label{rel:compat-a}
\end{align}
Using the definition $\mb{R}^{-T}=1-a\ket{\varsigma_0}\bra{\mathfrak{f}(x,t)}$ and the relation $\partial_t \bra{\mathfrak{f}(x,t)} +\partial_{x}\bra{\mathfrak{f}(x,t)}\mb{A}^T=0$, the above relation can be proved as follows:
\begin{align}
\partial_{x}\ket{\mf{g}^t_n(x,t)} - \partial_t\ket{\mf{g}^x_n(x,t)} &= \left(\partial_{x} \mb{R}^{-T}\right) \left( \partial_t \ket{\mf{h}(x,t)}\right) - \left(\partial_t \mb{R}^{-T}\right) \left( \partial_{x} \ket{\mf{h}(x,t)}\right) \cr
&=-a\ket{\varsigma_0}\left[\left(\partial_{x} \bra{\mathfrak{f}(x,t)}\right) \left( \partial_t \ket{\mf{h}(x,t)}\right) - \left(\partial_t \bra{\mathfrak{f}(x,t)}\right) \left( \partial_{x} \ket{\mf{h}(x,t)}\right) \right]\cr
&=a\ket{\varsigma_0}\underbrace{\left[\partial_{x} \bra{\mathfrak{f}(x,t)}\mb{A}^{-T}  +\partial_t \bra{\mathfrak{f}(x,t)} \right]}_{0}\left( \partial_{x} \ket{\mf{h}(x,t)}\right) \cr
&=0
\end{align}
The compatibility condition  in Eq.~\eqref{rel:compat-a} suggests that there exists a vector $\ket{\mf{g}_n(x,t)}$ such that 
\begin{align}
\mb{R}^{-T}\left(\partial_t \ket{\mf{h}(x,t)}\right) &= \ket{\mf{g}^t_n(x,t)} = \partial_t \ket{\mf{g}_n(x,t)}\\
\mb{R}^{-T}\left(\partial_{x} \ket{\mf{h}(x,t)}\right) &= \ket{\mf{g}^x_n(x,t)} = \partial_{x} \ket{\mf{g}_n(x,t)}. \label{def:p_xg_n}
\end{align}
In terms of this new vector Eq.~\eqref{eq:h-dig-1-a} can be re-written as 
\begin{align}
\partial_t \ket{\mf{g}_n(x,t)} + \mb{V}_{\rm eff} \partial_{x} \ket{\mf{g}_n(x,t)} =  \lambda \mb{R}^{-T}\ket{{\msc{A}(x},t)}, 
\label{eq:vec-g_n(x,t)-a}
\end{align}
or equivalently, componentwise 
\begin{align}
\partial_t \mf{g}_n(x,v,t) + v_{\rm eff}(x,v,t) \partial_{x} \mf{g}_n(x,v,t) = \lambda \bra{v}\mb{R}^{-T}\ket{{\msc{A}(x},t)}= \lambda({\msc{A}(x},v,t) - a \int du ~\mf{f}(x,u,t){\msc{A}(x},u,t)). \label{eq:g_n(x,t)-a}
\end{align}

\subsection{Finding $\mff(x,v,0)$ }
\label{find_df_r(0)-hr}
Since we are interested to find two-point correlation, as mentioned earlier, it is enough to find the solution of the saddle-point equations \eqref{eq:saddle-a} to linear order in $\lambda$. 
Inserting  $\mf{f}(x,v,t) \simeq \bmf{f}(x,v,t)+\lambda \delta \mf{f}(x,v,t)$,~$\mf{h}(x,v,t) \simeq \bmf{h}(x,v,t)+\lambda \delta \mf{h}(x,v,t)$ in Eq.~\eqref{eq:saddle-a}, one requires to solve them order by order in $\lambda$. The correlation function can then be obtained by inserting the final solution in Eq.~\eqref{def:den-corr-2}. 

To proceed, we first notice that in this case ${\msc{A}(x},v,t)= \msc{B}(x,v)\delta(t-t_a)$ with $\msc{B}(x,v)=\delta(v-u_a) \delta(x-x_a)$.
 This implies $\bmf{h}(x,v,t)=\bmf{g}_n(x,v,t)=0,~\forall~t$ and 
$\delta \mf{h}(x,v,T)=0$ for some $T>t_a$. Hence the Eq.~\eqref{eq:sad-1-a} becomes
\begin{align}
&\partial_t \delta \mf{h}_{\rm sd}(x,v,t)+\int du~ \mb{A}_u^v\partial_{x}\delta \mf{h}_{\rm sd}(x,u,t) = \msc{B}(x,v)\delta(t-t_a),~~\text{with}~~\msc{B}(x,v)=\delta(x-x_a)\delta(v-u_a).
 \label{eq:sad-1-1-a}
\end{align} 
Similarly, Eq.~\eqref{eq:g_n(x,t)-a} becomes 
\begin{align}
\partial_t \delta\mf{g}_n(x,v,t) + v_{\rm eff}(x,v,t) \partial_{x} \delta \mf{g}_n(x,v,t) =0, ~~\text{for}~0\leq t \le t_a. \label{eq:vec-dg(x,t)-a}
\end{align}
This equation can be solved easily by defining  the function $g(x',v,t)$ such that 
\begin{align}
{\delta \mf{g}_n(x,v,t)} = g(x-a\msc{F}_r(x,t),v,t), \label{def:psi^0-a}
\end{align}
and $g(x',v,t)$ satisfies the simple equation $\partial_tg(x',v,t)+v\partial_{x'}g(x',v,t)=0$. Given initial condition $g(x',v,0)$ the solution of this equation can be simply written as $g(x',v,t)=g(x'-vt,v,0)$.
Integrating Eq.~\eqref{eq:sad-1-1-a} over $t$ from $t_a -\epsilon$ to $t_a + \epsilon$, we get 
\begin{align}
\delta \mf{h}(x,v,t_a^-) = - \msc{B}(x,v),
\end{align}
which, from Eq.~\eqref{def:p_xg_n},  immediately implies
\begin{align}
\delta \mf{g}_n(x,v,t_a) &=\delta \mf{g}_n(-\infty,v,t_a)+ \int_{-\infty}^x dy~\bra{v}\mb{R}^{-T} \partial_y \ket{\delta \mf{h}(y,t_a^-)} \cr
&=\delta \mf{g}_n(-\infty,v,t_a)+\int_{-\infty}^x dy \bra{v}\left( 1-a\ket{\varsigma_0}\bra{\mf{f}(t_a)}\right)\partial_y \ket{\delta \mf{h}(y,t_a^-)}\cr 
& =-\hmc{R}_{\bmf{f}(t_a)}[\msc{B}](x,v),  \label{def:dmfg_n(t_a)}
\end{align}
where the operation $\hmc{R}_{\mf{f}}$ is defined as 
\begin{align}
\hmc{R}_{\mff(t)}[H(x,v)] = H(x,v) -a \int dy \int du~\Theta(x-y)~\mff(y,u,t)\partial_y H(y,u). \label{hR[H]} 
\end{align}
In the last line of  Eq.~\eqref{def:dmfg_n(t_a)} we have put $\delta \mf{g}_n(-\infty,v,t_a)=0$, since we do not expect any fluctuation at $x=-\infty$ to contribute to the two-point correlation between densities at two finite locations.
Inserting the explicit expression $\msc{B}(x,v)=\delta(x-x_a)\delta(v-u_a)$, one can simplify and get 
\begin{align}
\delta \mf{g}_n(x,v,t_a)
&= -\left(\msc{B}(x,v) -a \int dy \int du~\Theta(x-y)~\bmf{f}(y,u,t_a)\partial_y \msc{B}(y,u) \right) \cr
&= -\left(\msc{B}(x,v) +a \int dy~ \partial_y\left[\Theta(x-y)~\bmf{f}(y,u_a,t_a)\right]  \delta(y-x_a) \right) \cr
&= -\left(\msc{B}(x,v) +a\partial_{x_a} \left[ \Theta(x-x_a)~\bmf{f}(x_a,u_a,t_a)\right] \right) 
\label{mfg_n-t_a}
\end{align}
Given the value of $\delta \mf{g}_n(x,v,t)$ at $t=t_a$, we evolve Eq.~\eqref{eq:vec-dg(x,t)-a} backward in time from $t_a$ to $t=0$. For this we use the transformation in Eq.~\eqref{def:psi^0-a} and evolve the equation for $\delta \mf{g}_n$ backward in time. We finally get 
\begin{align}
\delta \mf{g}_n(x,v,0) = \delta g(x'_{\bmf{f}(0)}(x),v,0)&=\delta g(x'_{\bmf{f}(0)}(x)+vt_a,v,t_a) = \delta \mf{g}_n(x_{\bar{\fn}(t_a)}(x'_{\bmf{f}(0)}(x)+vt_a),v,t_a), \label{eq:dmtg_n(x,v,0)}
\end{align}
Inserting $\delta \mf{g}_n(x,v,t_a) $ from Eq.~\eqref{mfg_n-t_a}, we get 
\begin{align}
\delta \mf{g}_n(x,v,0)&= -\hmc{R}_{\bmf{f}(t_a)}[\msc{B}(x_{\bar{\fn}(t_a)}(x'_{\bmf{f}(0)}(x)+vt_a),v)].~\label{bkw-ev-mfg_n}
\end{align}
Once again performing the same procedure in Eq.~\eqref{def:p_xg_n} but now at $t=0$, we get \\ 
$\delta \mf{h}(x,v,0) = \int_{-\infty}^xdy~\bra{v}\mb{R}^T\partial_y \ket{\delta \mf{g}_n(y,0)}$.
Using the definition of $\mb{R}^T$ from Eq.~\eqref{eq:R^T-Rinv^T} we write 
\begin{align}
\delta \mf{h}(x,v,0) 
&=  \hmsc{R}_{\bmf{f}(0)}[ \delta \mf{g}_n(x,v,0)]=\hmsc{R}_{\bmf{f}(0)}[ \delta \mf{g}_n(x_{\bar{\fn}(t_a)}(x'_{\bmf{f}(0)}(x)+vt_a),v,t_a)], 
\label{dmfh(0)}
\end{align}
where $ \delta \mf{g}_n(x,v,0)$ is given in Eq.~\eqref{eq:dmtg_n(x,v,0)} and  we define the operation $\hmsc{R}_{\mf{f}(0)}$ on some function $H$ as
\begin{align}
\hmsc{R}_{\mff(t)}[H(x,v)]=H(x,v)+a \int dy~\Theta(x-y)\int du \frac{\mff(y,u,t)}{1-a\brvr(y,t)}\partial_y H(y,u). \label{hmscR[H]}
\end{align}
Note in order to obtain Eq.~\eqref{dmfh(0)}, we have put 
$ \delta \mf{h}(-\infty,v,0) =0$ 
due to the same reason as $\delta \mf{g}_n(-\infty,v,t_a)=0$. {One can follow a same procedure as in sec.~\ref{derv:f(x,v,0)} to show 
\begin{align}
\hmsc{R}_{\mff(t)}[\hmc{R}_{\mff(t)}[H(x,v)]]=H(x,v), \label{hmscRhmcR=I}
\end{align}
which says the operation $\hmsc{R}_{\mff(t)}$ is inverse of the operator $\hmc{R}_{\mff(t)}$.}

Our next plan is to first  find $\mf{f}(x,v,0)$ to linear order in $\lambda$ from 
Eq.~\eqref{eq:sad-4-a} with $\mf{h}(x,v,0)=\lambda \delta \mf{h}(x,v,0)$. After that, we evolve $\mf{f}(x,v,0)$ to time $t_b$ to obtain 
$\mf{f}(x,v,t_b)\simeq\bmf{f}(x,v,t_b)+\lambda \delta \mf{f}(x,v,t_b)$, inserting which in Eq.~\eqref{def:den-corr-2} will finally provide the two-point correlation. 
Solving Eq.~\eqref{eq:sad-4-a}, for given $\mf{h}(x,v,0)$, we get the same equation as in \eqref{eq:bmf(0)} for $\bmf{f}(x,v,0)$ (as expected) and for, $\delta \mf{f}(x,v,0)$ we get
\begin{align}
\delta \mf{f}(x,v,0)= \bmf{f}(x,v,0)\left[ \delta \mf{h}(x,v,0) -a(2-a\brvr(x,0) \int du~\bmf{f}(x,u,0)\delta \mf{h}(x,u,0)\right], \label{dmf(f)-a}
\end{align}
as we obtained earlier from point particle approach in Eq.~\eqref{sol:dbmf(0)-fn} with $\delta \mt{h}_{t_a}(x,v)$ replaced by $\delta \mf{h}(x,v,0)$.  It is indeed possible to show that $\delta \mt{h}_{t_a}(x,v)=\delta \mf{h}(x,v,0)$ [see Eq.~\eqref{eq:h_ta=-mfh} in~\ref{h_ta=-mfh}]. Hence Eq.~\eqref{dmf(f)-a} is exactly same as Eq.~\eqref{sol:dbmf(0)-fn}. One now needs to evolve this deviation from $t=0$ to $t=t_b$ and that can be done exactly the same way as done in sec.~\ref{sec:evo-bmf(f)} which is essentially same as the method discussed in \cite{doyon2023ballistic}.

\section{Proof of $\delta \mt{h}_{t_a}(x,v)=\delta \mf{h}(x,v,0)$}
\label{h_ta=-mfh}
We first recall $\delta \mt{h}_{t_a}(x,v)$ from Eq.~\eqref{def:mth-0}
\begin{align}
\delta \mt{h}_{t_a}(x,v)&= \mathscr{R}_{\bmf{f}(0)} \left[\delta \mt{g}_0( x'_{\bmf{f}(0)}(x),v) \right]
=\mathscr{R}_{\bmf{f}(0)}[\delta \mt{g}_{t_a}(x_{\brf(t_a)}(x'_{\bmf{f}(0)}(x)+vt_a),v)], \cr 
&= -\mathscr{R}_{\bmf{f}(0)} \left[\mathcal{R}_{\bmf{f}(t_a)}[\msc{B}(x_{{\brf}(t_a)}(x'_{\bmf{f}(0)}(x)+vt_a),v)] \right], \label{def:mth-a}
\end{align}
where the operations $\mc{R}_{\bmf{f}(t)}$ and $\msc{R}_{\bmf{f}(0)}$ are defined in Eq.~\eqref{R[H]} and Eq.~\eqref{Rsc[H]}, respectively. It is easy to show that 
\begin{align}
\mc{R}_{\bmf{f}(t_a)}[H(x,v)]
&=H(x,v) + a \int dy \int du ~\Theta\big{(}y-x\big{)}\left\{ \partial_{y} H(y,u)\right\}\bmf{f}(y,u,t_a), \cr
&= \hmc{R}_{\bmf{f}(t_a)}[H(x,v) ] +a \msc{K}_H(t_a) \label{mcR-hmcR}
\end{align}
where 
\begin{align}
\msc{K}_H(t_a) = \int dy \int du~\left\{ \partial_{y} H(y,u)\right\}\bmf{f}(y,u,t_a), \label{mscK}
\end{align}
is a constant and $\hmc{R}_{f(0)}$ is defined in Eq.~\eqref{hR[H]}. 
Similarly, one can show that,
\begin{align}
\msc{R}_{\bmf{f}(0)}[H(x,v)]&=H(x,v) -a \int dy~\Theta(y-x)~\left[ \int du \frac{\bmf{f}(y,u,0)}{1-a\brvr(y,0)} \left\{ \partial_{y} H(y,u)\right\}\right], \cr
&=\hmsc{R}_{\bmf{f}(0)}[H(x,v)] -a  \msc{J}_H(0), \label{mscR-hmscR}
\end{align}
where 
\begin{align}
\msc{J}_H(0)&=  \int dy~ \int du \frac{\bmf{f}(y,u,0)}{1-a\brvr(y,0)} \left\{ \partial_{y} H(y,u)\right\},\cr
&= \int dy'~ \int du ~\bar{\fn}(y',u,0) \left\{ \partial_{y'} H(x_{{\brf}(0)}(y'),u)\right\},
\end{align}
is again a constant and recall the form of $\hmsc{R}_{\mf{f}(0)}$ from Eq.~\eqref{hmscR[H]}. 

Applying the relations from Eq.~\eqref{mcR-hmcR} and Eq.~\eqref{mscR-hmscR} on the right-hand side of Eq.~\eqref{def:mth-a}, we get
\begin{align}
\delta \mt{h}_{t_a}(x,v) &=-\msc{R}_{\bmf{f}(0)}[\mc{R}_{{\bmf{f}}(t_a)}[\msc{B}(x_{{\brf}(t_a)}(x'_{\bmf{f}(0)}(x)+vt_a),v)]]  \cr
&= -\msc{R}_{{\bmf{f}}(0)} \left[\hmc{R}_{{\bmf{f}}(t_a)}[\msc{B}(x_{{\brf}(t_a)}(x'_{\bmf{f}(0)}(x)+vt_a),v)] + a \msc{K}_{\msc{B}}(t_a)\right],\cr
&=- \msc{R}_{{\bmf{f}}(0)} \left[\hmc{R}_{{\bmf{f}}(t_a)}[\msc{B}(x_{{\brf}(t_a)}(x'_{\bmf{f}(0)}(x)+vt_a),v)]\right] - a \msc{K}_{\msc{B}}(t_a),\cr
&=- \hmsc{R}_{{\bmf{f}}(0)} \left[\hmc{R}_{{\bmf{f}}(t_a)}[\msc{B}(x_{{\brf}(t_a)}(x'_{\bmf{f}(0)}(x)+vt_a),v)]\right] +a\msc{J}_{\msc{B}}(0)- a \msc{K}_{\msc{B}}(t_a),\label{mth-a-1}
\end{align}
where 

\begin{align}
\msc{J}_{\msc{B}}(0)&=\int dy~ \int du \frac{\bmf{f}(y,u,0)}{1-a\brvr(y,0)} \left\{ \partial_{y} \hmc{R}_{\bmf{f}(t_a)}[\msc{B}(x_{{\brf}(t_a)}(x'_{\bmf{f}(0)}(y)+ut_a),u)]\right\}, \cr
&=\int dy~ \int du \frac{\bmf{f}(y,u,0)}{1-a\brvr(y,0)} \left\{ \partial_{y} \mc{R}_{\bmf{f}(t_a)}[\msc{B}(x_{{\brf}(t_a)}(x'_{\bmf{f}(0)}(y)+ut_a),u)]\right\}, \cr
&=\int dy'~ \int du~\bar{\fn}(y',u,0) \left\{ \partial_{y'} \mc{R}_{\bmf{f}(t_a)}[\msc{B}(x_{{\brf}(t_a)}(y'+ut_a),u)]\right\}, \label{J_A-1}
\end{align}
where on the second line we have again used $\hmc{R}_{\bmf{f}(t_a)}=\mc{R}_{\bmf{f}(t_a)}-a\msc{K}_{\msc{B}}$. Now, inserting the explicit expression of $\mc{R}_{\bmf{f}(t_a)}[\msc{B}(x_{{\brf}(t_a)}(y'+ut_a),u)]$ from Eq.~\eqref{R[H]} into Eq.~\eqref{J_A-1}, we get the following 
\begin{align}
\msc{J}_{\msc{B}}(0)
&=\int dy'~ \int du~\bar{\fn}(y',u,0) \partial_{y'}\big{[}\msc{B}(x_{{\brf}(t_a)}(y'+ut_a),u)\big{]} \cr 
&~~-a \int dy'~ \int du~\bar{\fn}(y',u,0) \partial_{y'} \left [ \int dz' \int dw~\partial_{x_{{\brf}(t_a)}(z')}\left\{\msc{B}(x_{{\brf}(t_a)}(z'),w)\right\}\Theta(y'+ut_a-z')\bar{\fn}(z',w,t_a)\right ], \cr
&=\int dz'~ \int du~\bar{\fn}(z'-ut_a,u,0) \partial_{z'}\big{[}\msc{B}(x_{{\brf}(t_a)}(z'),u)\big{]} \cr 
&~~-a \int dy'~ \int du~\bar{\fn}(y',u,0) \left [ \int dz' \int dw~\partial_{x_{{\brf}(t_a)}(z')}\left\{\msc{B}(x_{{\brf}(t_a)}(z'),w)\right\}\delta(y'+ut_a-z')\bar{\fn}(z',w,t_a)\right ], \cr
&=\int dz'~ \int du~\bar{\fn}(z',u,t_a) \partial_{z'}\big{[}\msc{B}(x_{{\brf}(t_a)}(z'),u)\big{]} \cr 
&~~-a \int dz'~ \int du~\bar{\fn}(z'-ut_a,u,0) \left [  \int dw~\partial_{x_{{\brf}(t_a)}(z')}\left\{\msc{B}(x_{{\brf}(t_a)}(z'),w)\right\}\bar{\fn}(z',w,t_a)\right ], \cr
&=\int dz'~ \int du~\bar{\fn}(z',u,t_a) \partial_{x_{{{\brf}}(t_a)}(z')}\big{[}\msc{B}(x_{{\brf}(t_a)}(z'),u)\big{]} (1+a \brr(z',t_a))\cr 
&~~-a \int dz'~ \int dw~ \brr(z',t_a) \left [ \partial_{x_{{{\brf}(t_a)}}(z')}\left\{\msc{B}(x_{{\brf}(t_a)}(z'),w)\right\}\bar{\fn}(z',w,t_a)\right ], \cr
&=\int dz'~ \int du~\bar{\fn}(z',u,t_a) \partial_{x_{{{\brf}(t_a)}}(z')}\big{[}\msc{B}(x_{{\brf}(t_a)}(z'),u)\big{]}, \cr
&=\int dz~ \int du~\bmf{f}(z,u,t_a) \partial_{z}\big{[}\msc{B}(z,u)\big{]} = \msc{K}_{\msc{B}}(t_a)
\end{align}
Hence, the two constant terms in Eq.~\eqref{mth-a-1} perfectly cancel each other, and we have 
\begin{align}
\delta \mt{h}_{t_a}(x,v) =-\hmsc{R}_{\bmf{f}(0)} \left[\hmc{R}_{\bmf{f}(t_a)}[\msc{B}(x_{{\brf}(t_a)}(x'_{\bmf{f}(0)}(x)+vt_a),v)]\right] = \delta \mf{h}(x,v,0). \label{eq:h_ta=-mfh}
\end{align}
according to Eq.~\eqref{dmfh(0)}.

\section{Derivation of Eq.~\eqref{eq:dmscf(t)}}
\label{derv:df_r}
We start with Eq.~\eqref{eq:f_f^lam(t)-->f_r^lam(t)-mt}, which, for convenience, we here rewrite
\begin{align}
\mff_\lambda(z,v,t) = \frac{\fn_\lambda(x'_{\mff_\lambda(t)}(z),v,t)}{1+a\rho^\lambda(x'_{\mff_\lambda(t)}(z),t)},~~~\text{with}~~x'_{\mff_\lambda(t)}(z) = z- a \msc{F}_{\rm r}^\lambda(z,t). \label{eq:f_f^lam(t)-->f_r^lam(t)}
\end{align}
Taking the derivative with respect to $\lambda$ on both sides of Eq.~\eqref{eq:f_f^lam(t)-->f_r^lam(t)} and evaluating them at $\lambda=0$ one finds
\begin{align}
\delta \mff(z,v,t) &= \frac{ {\delta \fn(}x'_{\bmf{f}(t)}(z),v,t)}{1+a\brr(x'_{\bmf{f}(t)}(z),t)} - a \frac{\bar{\fn}(x'_{\bmf{f}(t)}(z),v,t)~\delta \rho(x'_{\bmf{f}(t)}(z),t)}{(1+a\brr(x'_{\bmf{f}(t)}(z),t))^2}~-~a \frac{\delta \msc{F}_{\rm r}(z,t)}{1-a\brvr(z,t)}\partial_z\bmf{f}(z,v,t).
\label{eq:dmscf(t)-0}
\end{align}
where 
\begin{align}
\delta \rho(z',t)= \int du~ {\delta \fn(}z',u,t) = \int du~\left[ \partial_\lambda \fn_\lambda(z',u,t)\right]_{\lambda=0},
\end{align}
and 
\begin{align}
\delta \msc{F}_{\rm r}(z,t) =\int dy\Theta(z-y) \int du~\delta \mff(z,u,t).
\end{align}
Note that Eq.~\eqref{eq:dmscf(t)-0} contains two unknowns, $\delta \mff(z,v,t)$ and $\delta \msc{F}_{\rm r}(z,t)$ on both sides of it. Performing integration on $v$, one can eliminate $\delta \msc{F}_{\rm r}(z,t)$ as follows. One gets 
\begin{align}
\delta\varrho(z,t) = \int du~\delta \mff(z.u,t)= \frac{\delta \rho(x'_{\bmf{f}(t)}(z),t)}{(1+a\brr(x'_{\bmf{f}(t)}(z),t))^2} - a \delta \msc{F}_{\rm r}(z,t)\frac{\partial_z \brvr(z,t)}{1-a\brvr(z,t)}. \label{def:vrho(t)-1}
\end{align}
Noting $\delta F(z',t)= \int_{-\infty}^{z'}dy'~\delta \rho(y',t)$, we can rewrite the above equation as 
\begin{align}
\partial_z\left[ \delta F(x'_{\bmf{f}(t)}(z),t)-   \frac{\delta \msc{F}_{\rm r}(z,t)}{1-a\brvr(z,t)} \right]=0,
\end{align}
which upon integration yields 
\begin{align}
 \delta F(x'_{\bmf{f}(t)}(z),t) =   \frac{\delta \msc{F}_{\rm r}(z,t)}{1-a\brvr(z,t)}, \label{rel:dF-dmfF}
\end{align}
where we have assumed $\delta \msc{F}_{\rm r}(z \to -\infty,t) =0$, $ x'_{\bmf{f}(t)}(z \to -\infty) \to -\infty$ and $\delta {F}(z' \to -\infty,t) =0$. The relation in Eq.~\eqref{rel:dF-dmfF} is generically true and can be easily proved from a more general consideration. Using Eq.~\eqref{rel:dF-dmfF}  in Eq.~\eqref{eq:dmscf(t)-0}, we get 
\begin{align}
\delta \mff(z,v,t) &= \frac{ {\delta \fn(}x'_{\bmf{f}(t)}(z),v,t)}{1+a\brr(x'_{\bmf{f}(t)}(z),t)} - a \frac{\bar{\fn}(x'_{\bmf{f}(t)}(z),v,t)~\delta \rho(x'_{\bmf{f}(t)}(z),t)}{(1+a\brr(x'_{\bmf{f}(t)}(z),t))^2}~-~a \delta F(x'_{\bmf{f}(t)}(z),t) \partial_z\bmf{f}(z,v,t),
\label{eq:dmscf(t)-1-a}
\end{align}
where, according to Eq.~\eqref{sol:f(t)} we have $\bar{\fn}(z',v,t)=\bar{\fn}(z'-vt,v,0)$ and $ {\delta \fn(}z',v,t)= {\delta \fn(}z'-vt,v,0)$. Since  ${\delta \fn(}z',v,0) = (\partial_\lambda \fn_\lambda(z',v,0))_{\lambda=0}$  can be determined from Eq.~\eqref{eq:f_r^lam(0)->f^lam(0)} {\it i.e.}, essentially from $\delta \mff(z,v,0)$ in Eq.~\eqref{sol:dbmf(0)-fn}, the right-hand side of Eq.~\eqref{eq:dmscf(t)} completely determines $\delta \mff(z,v,t) $ at any phase space point $(z,v)$ at time $t$ for given $\bmf{f}(z,v,0)$. Once again, using Eq.~\eqref{rel:dF-dmfF} in Eq.~\eqref{def:vrho(t)-1} one can simplify the expression of $\delta\varrho(z,t) $ and get 
\begin{align}
\delta \varrho(z,t) 
&=\partial_z\left[ (1-a\brvr(z,t))\delta\bar{F}(x'_{\bmf{f}(t)}(z),t)\right]. \label{eq:dbrvr(z,t)-fn}
\end{align}
In the next step, we need to find ${\delta \fn(}z',v,t)$, $\delta \rho(z',t)$ and $\delta F(z',t)$. 
Taking the derivative with respect to $\lambda$ on both sides of Eq.~\eqref{eq:f_r^lam(0)->f^lam(0)} at $t=0$ we get  
\begin{align}
\begin{split}
 {\delta \fn(}x',v,0)= \frac{\delta \mff(x_{{\brf}(0)}(x'),v,0)}{1-a\brvr(x_{{\brf}(0)}(x'),0)}
 &+a\frac{\bmf{f}(x_{{\brf}(0)}(x'),v,0)~\delta \varrho(x_{{\brf}(0)}(x'),0)}{(1-a\brvr(x_{{\brf}(0)}(x'),0))^2} \\ 
 &+ a\delta \msc{F}_{\rm r}(x_{{\brf}(0)}(x'),0)\partial_{x'}~\frac{\bmf{f}(x_{{\brf}(0)}(x'),v,0)}{1-a\brvr(x_{{\brf}(0)}(x'),0)},
 \end{split}
 \label{ex:df(0)-1}
\end{align}
which can be used to compute ${\delta \fn(}z',v,t) = {\delta \fn(}z'-vt,v,0)$. Performing a few simplifications, we finally get  
\begin{align}
 {\delta \fn(}z',v,t) &= {\delta \fn(}z'-vt,v,0) \cr 
&=  \bar{\fn}(z'-vt,v,0)~\mt{h}_{t_a}^{\rm dr}(x_{{\brf}(0)}(z'-vt),v) + a [\partial_{z'}  \brf(z'-vt,v,0)]\delta \msc{F}_{\rm r}(x_{\brf(0)}(z'-vt),0)\cr 
&= \bar{\fn}(z',v,t)~\mt{h}_{t_a}^{\rm dr}(x_{{\brf}(0)}(z'-vt),v)+ a [\partial_{z'}  \brf(z',v,t)]\delta \msc{F}_{\rm r}(x_{\brf(0)}(z'-vt),0), \label{eq:df(t)}
\end{align}
where 
\begin{align}
\delta \msc{F}_{\rm r}(x,0)=\int dy~\Theta(x-y)~\delta \varrho(x,0)=\int dy~\Theta(x-y)~(1-a\brvr(y,0))^2~\int du~\bmf{f}(y,u,0)~\delta \mt{h}_{t_a}(y,u). \label{def:dbmfF(0)}
\end{align}
  with $\delta h_{t_a}(z,v)$ given in Eq.~\eqref{def:mth-0} and
  \begin{align}
\delta \mt{h}_{t_a}^{\rm dr}(x,v) =& \delta \mt{h}_{t_a}(x,v) - a\int du~\bmf{f}(x,u,0)~\delta \mt{h}_{t_a}(x,u). \label{mth_t_a^dr-a-mt}
 \end{align}
Integrating over $v$, one gets
\begin{align}
\delta \rho(z',t)&= \int du~ {\delta \fn(}z',u,t)\cr 
&=  \int du~\brf(z',u,t)~\mt{h}_{t_a}^{\rm dr}(x_{{\brf}(0)}(z'-ut),u) + a \int du~[\partial_{z'}  \brf(z',u,t)]\delta \msc{F}_{\rm r}(x_{\brf(0)}(z'-ut),0). 
\label{def:drho(t)}
\end{align}
Similarly, integrating $\delta \rho(y',t)$ one gets 
\begin{align}
\delta F(z',t)&= \int_{-\infty}^{z'}dy'~\delta \rho(y',t) \cr 
&= \int dy'\Theta(z'-y')\int du~\brf(y',u,t)~\delta\mt{h}_{t_a}^{\rm dr}(x_{{\brf}(0)}(y'-ut),u), \cr
&~~~~~~~~~~~~~
+a\int dy'\Theta(z'-y') \int du~[\partial_{y'}  \brf(y',u,t)]\delta \msc{F}_{\rm r}(x_{\brf(0)}(y'-ut),0), \cr
&= \int dy~\Theta(x_{\brf(t)}(z')-y)\int du~\bmf{f}(y,u,t)~\mt{h}_{t_a}^{\rm dr}(x_{{\brf}(0)}(x'_{\bmf{f}(t)}(y)-ut),u) \cr 
&~~~~~~~~~~~~~
+a\int dy\Theta(x_{\brf(t)}(z')-y) \int du~\partial_{y} \left( \frac{\bmf{f}(y,u,t)}{1-a\brvr(y,t)}\right)\delta \msc{F}_{\rm r}(x_{\brf(0)}(x'_{\bmf{f}(t)}(y)-ut),0).
 \label{def:dF(t)-a}
\end{align}
Inserting the forms of $ {\delta \fn(}z',v,t)$, $\delta \rho(z',t)$ and $\delta F(z',t)$ from Eqs.~(\ref{eq:df(t)}), (\ref{def:drho(t)}) and \eqref{def:dF(t)-a}, respectively, in {Eq.~\eqref{eq:dmscf(t)-1-a}} and simplifying we get 
\begin{align}
\delta \mff(z,v,t) &=\bmf{f}(z,v,t)\left[\delta \mt{h}_{t_a}^{\rm dr}(x_{{\brf}(0)}(x'_{\bmf{f}(t)}(z)-vt),v)-a\int dw~\bmf{f}(z,w,t)~\delta \mt{h}_{t_a}^{\rm dr}(x_{{\brf}(0)}(x'_{\bmf{f}(t)}(z)-wt),w) \right]   \cr 
&~~~~~~~+a\Big{\{}\partial_{z} \left[ \frac{\bmf{f}(z,v,t)}{1-a\brvr(z,t)}\right]\delta \msc{F}_{\rm r}(x_{\brf(0)}(x'_{\bmf{f}(t)}(z)-vt),0)   \displaybreak[3]  \\
&-a \bmf{f}(z,v,t) \int dw~\partial_{z} \left[ \frac{\bmf{f}(z,w,t)}{1-a\brvr(z,t)}\right]\delta \msc{F}_{\rm r}(x_{\brf(0)}(x'_{\bmf{f}(t)}(z)-wt),0)\Big{\}} 
-a \delta F(x'_{\bmf{f}(t)}(z),t)\partial_z\bmf{f}(z,v,t), \displaybreak[3] \cr 
&=\bmf{f}(z,v,t)\delta \mt{h}_{t_a}^{\rm dr}(x_{{\brf}(0)}(x'_{\bmf{f}(t)}(z)-vt),v) - a \partial_z \left[ \bmf{f}(z,v,t) \delta F(x'_{\bmf{f}(t)}(z),t)\right] \displaybreak[3] \cr 
&~~~~~~~~~~~~~~~~~~~~~~~~~~
+a \partial_{z} \left[ \frac{\bmf{f}(z,v,t)}{1-a\brvr(z,t)}\right]\delta \msc{F}_{\rm r}(x_{\brf(0)}(x'_{\bmf{f}(t)}(z)-vt),0). \displaybreak[3] \label{eq:dmscf(t)-a}
\end{align}
Integrating both sides of the above equation over $v$ we get 
\begin{align}
\delta \varrho(z,t) 
&=(1-a\brvr(z,t)) \Big{\{}\int dw~\bmf{f}(z,w,t)~\delta \mt{h}_{t_a}^{\rm dr}(x_{{\brf}(0)}(x'_{\bmf{f}(t)}(z)-wt),w) \cr 
&~~~~~~
+a\int dw~\partial_{z} \left[ \frac{\bmf{f}(z,w,t)}{1-a\brvr(z,t)}\right]\delta \msc{F}_{\rm r}(x_{\brf(0)}(x'_{\bmf{f}(t)}(z)-wt),0)  - a \frac{\delta F(x'_{\bmf{f}(t)}(z),t)\partial_z\brvr(z,t)}{1-a\brvr(z,t)}  \Big{\}}, 
\label{eq:dbrvr(z,t)-fn-1-a}
\end{align}
which also expectedly reduces to the expression in Eq.~\eqref{eq:dbrvr(z,t)-fn}. For $t=0$, it is easy to see by using Eq.~\eqref{rel:dF-dmfF} that the above expression correctly reduces to the expression of $\delta \mff(z,v,0)$ in Eq.~\eqref{sol:dbmf(0)-fn}.

\section{Some details on the function $\delta \mt{h}_{t_a}^{\rm dr}(x,v)$}
\label{sec:dh_t_a^dr}
We want to compute $\delta \mt{h}_{t_a}^{\rm dr}(x,v)$ defined in Eq.~\eqref{mth_t_a^dr} which we rewrite here for convenience
\begin{align}
\delta \mt{h}_{t_a}^{\rm dr}(x,v)=& \delta \mt{h}_{t_a}(x,v) - a\int du~\bmf{f}(x,u,0)~\delta \mt{h}_{t_a}(x,u), \label{mth_t_a^dr-a}
\end{align}
To evaluate $\delta \mt{h}_{t_a}(x,v)$ we refer to Eq.~\eqref{def:mth-0}. Denoting $\delta \mt{g}_n(x,v,0)=\delta \mt{g}_{0}(x'_{\bmf{f}(0)}(x),v)$ and using Eq.~\eqref{Rsc[H]}, we write 
\begin{align}
\delta \mt{h}_{t_a}(x,v)&=\delta \mt{g}_{n}(x,v,0) - a \int dy~ \Theta \left(y-x \right) \int dw ~\frac{\bmf{f}(y,w,0)}{1-a\brvr(y,0)}\partial_y \delta \mt{g}_{n}(y,w,0),
\label{eq:mth-a-3}
\end{align}
Inserting the expression from Eq.~\eqref{eq:mth-a-3} in Eq.~\eqref{mth_t_a^dr-a}, we have
\begin{align}
\delta \mt{h}_{t_a}^{\rm dr}(x,v) 
&=\delta \mt{g}_{n}(x,v,0) -a \int dw~\bmf{f}(x,w,0)\delta\mt{g}_{n}(x,w,0)\cr
&~~~-a(1-a \brvr(x,0))\int dy  \int dw ~\frac{\bmf{f}(y,w,0)}{1-a\brvr(y,0)} \Theta(y-x)\partial_y\delta\mt{g}_{n}(y,w,0).
\label{mthdr-a-2}
\end{align}
where $\delta \mt{g}_n(x,v,0)=\delta \mt{g}_0(x'_{\bmf{f}(0)}(x),v)$ is given by Eq.~\eqref{def:mth-0} as  
\begin{align}
\delta \mt{g}_n(x,v,0)=\delta \mt{g}_0(x'_{\bmf{f}(0)}(x),v)=\delta \mt{g}_{t_a}(x_{\brf(t_a)}(x'_{\bmf{f}(0)}(x)+vt_a),v), \label{rel:mtg(0)-mtg(t_a)}
\end{align}
with [see Eq.~\eqref{def:mtg_t_a}]
\begin{align}
 \delta \mt{g}_{t_a}(y,w)&= -\mathcal{R}_{\bmf{f}(t_a)}[\msc{B}(y,w)]=-\left[ \msc{B}(y,w) - a\partial_{x_a}[\Theta(x_a-y)\bmf{f}(x_a,u_a,t_a)] \right]. \label{def:mtg_t_a-a}
\end{align}
It is easy to see from Eq.~\eqref{mth_t_a^dr-a} that
\begin{align}
\int dw~\bmf{f}(x,w,0)~\delta \mt{h}_{t_a}^{\rm dr}(x,w)&= (1-a\brvr(x,0)) \int dw~\bmf{f}(x,w,0) \delta\mt{h}_{t_a}(x,w,0), \cr
\begin{split}
\text{where,}~~ \int dw~\bmf{f}(x,w,0) \delta\mt{h}_{t_a}(x,w,0)&=\bigg{[} \int dw~\bmf{f}(x,w,0)\delta\mt{g}_{n}(x,w,0)\cr
&~~-a\brvr(x,0) \int dy\int dw\frac{\bmf{f}(y,w,0)}{1-a\brvr(y,0)}\Theta(y-x)\partial_y\delta \mt{g}_{n}(y,w,0) \bigg{]}. 
\end{split}
\label{int-f_r(w)h^dr(w)}
\end{align}
\paragraph{In Equilibrium:} If the hard-rod gas initially is in equilibrium {\it i.e.}, $\bmf{f}(x,v,0)=\bmf{f}_{\rm eq}(x,v)=\brvr_{\rm eq}~\gn(v)$ then the PSD at later time also remains the same, which means $\bmf{f}(x,v,t)=\bmf{f}_{\rm eq}(x,v),~~\forall t$. Since $\bmf{f}_{\rm eq}(x,v)=\brvr_{\rm eq}~\gn(v)$ is independent of space, the expression of $\delta \mt{h}_{t_a}(x,v) $ in Eq.~\eqref{eq:mth-a-3} simplifies to 
\begin{align}
\delta \mt{h}_{t_a}(x,v)  &=\delta \mt{g}_{n}(x,v,0) +\frac{a \brvr_{\rm eq}}{1-a \brvr_{\rm eq}} \int dw~\gn(w)\delta\mt{g}_{n}(x,w,0),
\label{mthdr-a-2bis}
\end{align}
which implies
\begin{align}
\int dw~ \bmf{f}(x,w,0)\delta \mt{h}_{t_a}(x,w) = \frac{\brvr_{\rm eq}}{1-a \brvr_{\rm eq}} \int dw~\gn(w)\delta\mt{g}_{n}(x,w,0). \label{int-f_r(w)h(w)-eq}
\end{align}
Inserting this result in the expression of equilibrium space-time correlation in Eq.~\eqref{ex:corr-ue-t-1} and simplifying we get 
\begin{align}
\mc{S}_{\rm eq}(x_a,u_a,t_a;x_b,u_b,0)= -  \brvr_{\rm eq}\gn(u_b) \left[\delta \mt{g}_{n}(x_b,u_b,0)- a\brvr_{\rm eq}~\int dw~\gn(w)\delta\mt{g}_{n}(x_b,w,0) \right], \label{ex:corr-ue-t-1-a}
\end{align}
where, $\delta \mt{g}_n(x,v,0)$ given in Eq.~\eqref{rel:mtg(0)-mtg(t_a)} with  $ \delta \mt{g}_{t_a}(y,w)=-\delta(y-x_a)\left[\delta(w-u_a)  - a\brvr_{\rm eq}\gn(u_a) \right]$ in equilibrium. Inserting this form of $ \delta \mt{g}_{t_a}(y,w)$, we perform the integral. For that we first note that 
\begin{align}
\delta \left( x_{\brf(t_a)}(x'_{\bmf{f}(0)}(y)+wt_a)-x_a\right) = (1-a\brvr(x_a,t_a))~\delta \left(x'_{\bmf{f}(0)}(y)+wt_a -x'_{\bmf{f}(t_a)}(x_a)\right),
\end{align}
using which along with $\bmf{f}(x,v,0)=\bmf{f}_{\rm eq}(x,v)=\brvr_{\rm eq}~\gn(v)$ we get 
\begin{align}
\int dw~\gn(w)\delta\mt{g}_{n}(x_b,w,0)&= -(1-a\brvr_{\rm eq}) \frac{1}{t_a}\gn \left( \frac{\Delta x}{t_a}\right)~\left[ \delta  \left(u_a- \frac{\Delta x}{t_a}\right) -a \brvr_{\rm eq}\gn(u_a)\right], \\
&\text{where}~~~~\Delta x = x'_{\bmf{f}_{\rm eq}}(x_a)- x'_{\bmf{f}_{\rm eq}}(x_b) = (1-a\brvr_{\rm eq})~(x_a-x_b). \label{def:Dx}
\end{align}
Using the above expression of the integral in Eq.~\eqref{ex:corr-ue-t-1-a} we get the expression in Eq.~\eqref{ex:corr-ue-t-2}.

\section*{References}

\bibliographystyle{unsrt}
\bibliography{references}%

\end{document}